\documentclass[aps, prl, twocolumn, a4paper, 10pt, showpacs,reprint,superscriptaddress,nofootinbib]{revtex4-2}

\usepackage[left=2cm, right=2cm, top=2.5cm, bottom=2.5cm]{geometry}
\usepackage{graphicx} % Required for inserting images
\usepackage{amsmath}
\usepackage{braket}
\usepackage{bbm}
\usepackage{bm}
\usepackage{romannum}
\usepackage{color}
\usepackage[dvipsnames]{xcolor}
\usepackage{placeins}
\usepackage{hyperref}
\usepackage{orcidlink}
\usepackage{amsfonts} 
\usepackage{dsfont}

\renewcommand{\thesection}{\Roman{section}}
\renewcommand{\thesubsection}{\Alph{subsection}}

\definecolor{deeppurple}{rgb}{0.7, 0, 0.8}

\newcommand{\Ol}{\hat{O}_{\rm{L}}}
\newcommand{\Onl}{\hat{O}_{\rm{NL}}(\phi^{[j]})}
\newcommand{\expVal}{\braket{\sigma_z}_{\rm{anc}}}
\newcommand{\succProbs}{\alpha_{\rm{succ}}}
\newcommand{\phiOpt}{\varphi^{*}}

\begin{document}
\title{Efficient Treatment of Non-Linearity in Quantum Computational Fluid Dynamics Using Hybrid Tensor Networks}
\date{\today}
\author{Pia Siegl$^{\ast,\dag}$ \orcidlink{0000-0003-2249-8121}}
\affiliation{Institute of Software Methods for Product Virtualization, German Aerospace Center (DLR), Nöthnitzer Straße 46b, 01187 Dresden, Germany}
\affiliation{Institute for Quantum Physics, University of Hamburg, Luruper Chaussee 149, 22761 Hamburg, Germany}
\author{Nis-Luca van Hülst \orcidlink{0009-0004-9893-3614}}
\affiliation{Institute for Quantum Physics, University of Hamburg, Luruper Chaussee 149, 22761 Hamburg, Germany}
\author{Maximilian Mandelt Buxadé \orcidlink{0009-0006-5246-4415}}
\affiliation{Institute of Aerodynamics and Flow Technology, German Aerospace Center (DLR), Lilienthalplatz 7, 38108 Braunschweig, Germany}
\author{Tomohiro Hashizume \orcidlink{0000-0002-7154-5417}}
\affiliation{Institute for Quantum Physics, University of Hamburg, Luruper Chaussee 149, 22761 Hamburg, Germany}

\author{Dieter Jaksch \orcidlink{0000-0002-9704-3941}}
\affiliation{Institute for Quantum Physics, University of Hamburg, Luruper Chaussee 149, 22761 Hamburg, Germany}
\affiliation{The Hamburg Centre for Ultrafast Imaging, Hamburg, Germany}
\affiliation{Clarendon Laboratory, University of Oxford, Parks Road, Oxford OX1 3PU, UK}
\def\thefootnote{$\dag$}\footnotetext{Contact author: pia.siegl@dlr.de}

\begin{abstract}
%Nonlinear terms present a fundamental challenge for quantum computational fluid dynamics, as their implementation on inherently linear quantum hardware typically requires resource-intensive workarounds that limit scalability to large-scale simulations.
%We present a hybrid quantum-classical tensor network algorithm that tackles this challenge using a variational time stepping routine paired with quantum tensor programming to map operators and fields efficiently into quantum circuits. Relying on the probabilistic implementation of the non-linear operation, we substitute previous state-based implementations  with tensor-based block encodings to mitigate exponentially decreasing success probabilities.
%Bench marked at the example of turbulent flow fields, we show that the method succeeds with high success probability keeping the measurement requirements moderate even for increased system sizes. A comparison to classical tensor network mathods shows significant savings in computational cost and memory, leading the path for efficient quantum algorithms for demanding CFD simulations. 
Nonlinear terms present a fundamental challenge for quantum computational fluid dynamics, as their implementation on inherently linear quantum hardware typically requires resource-intensive workarounds that limit scalability to large-scale simulations. 
We present a hybrid quantum-classical tensor network algorithm that addresses this bottleneck by combining variational time-stepping with quantum tensor programming to efficiently compile operators and time-dependent fields into quantum circuits. Within a probabilistic framework, we replace prior state-based nonlinear implementations with tensor-based block encodings, stabilizing success probabilities that otherwise decay exponentially with system size. 
Benchmarking on turbulent flow fields demonstrates that the algorithm maintains high success probabilities and moderate measurement overhead across increasing Reynolds numbers and grid resolutions. Compared to fully classical tensor network solvers, our hybrid approach yields substantial reductions in both memory footprint and computational cost, establishing a scalable pathway toward practical quantum advantage in scale-resolving CFD simulations.
\end{abstract}

\maketitle
\pagenumbering{arabic}
\section{Introduction}
Computational fluid dynamics (CFD) enables critical advancements across aerospace, energy, climate science, and biomedical engineering \cite{WU2022}. Direct numerical simulation (DNS) is indispensable for capturing the multi-scale, highly nonlinear dynamics of turbulent flows. However, the computational cost of these methods scales steeply with the Reynolds number, rendering them intractable for many industrially and scientifically relevant regimes ~\cite{MOSER2023,Nasa2030,Moin2007}. This bottleneck stems from the rapid growth in degrees of freedom, which imposes severe memory and computational demands.

To mitigate these costs, quantum-inspired methods based on tensor networks, and quantum algorithmic approaches have attracted significant interest \cite{airbus_quantum_challenge}, promising substantial reductions in memory and computational overhead \cite{Gourianov2022,Ye2022, Kiffner2023, Kornev2023-arxiv,Ye2024, Peddinti2024, Hölscher2025, Gourianov2024, Siegl2026}. Their combination via quantum tensor programming offers a natural pathway to unite their strenghts:
recent advances enable systematic compilation of tensor network fields \cite{Ran2020} and operators \cite{Nibbi2024, Termanova2024} into sequences of unitary gates, with provable bounds on circuit depth.
The classical tensor train (TT) architecture \cite{Oseledets2011, Oseledets2013, Orus2014} represents fields and operators as chains of low-rank tensors, with compression efficiency governed by the bond dimension $\chi$. As a powerful compression and simulation tool, TTs have shown great promise for CFD, enabling significant memory savings and computational speedups \cite{Gourianov2022, Kiffner2023, Gourianov2024, Peddinti2024, vanHuelst2026}.
Quantum algorithms based on tensor programming leverage this compression strength and promise to provide additional polynomial speedups over its classical counterpart \cite{Lubasch2020, Siegl2026}, with increased potential gains from enhanced circuit expressivity \cite{Haghshenas2022, Siegl2026}. 

However, both paradigms face a critical bottleneck: the efficient implementation of nonlinearities, which are ubiquitous in CFD.
Although recent advancements in tensor network methods address the scaling of nonlinear operations \cite{Michailidis2025, Meng2026-arxiv}, they often inflate intermediate bond dimensions, negating asymptotic advantages.
Many quantum algorithms circumvent the linearity constraint of quantum evolution by embedding nonlinear dynamics into infinite-dimensional linear frameworks, such as Carleman linearization which enables compatibility with quantum linear solvers \cite{Krovi2023, Wang2026-arxiv} and Schrödingerization techniques \cite{Jin2024,Sasaki2026}. In practice, Carleman linearization requires finite-order truncation of the infinite-dimensional system of equations and recent analysis demonstrates that the required truncation order scales unfavorably with Reynolds number and grid resolution, introducing overhead that negates quantum advantages \cite{Gonzalez-Conde2025}.
%Truncation free linearization strategies can be exploited but these require extensive classical pre-processing \cite{Lacatus2026-arxiv}.
Alternative strategies such as quantum physics-informed neural networks \cite{Kyriienko2021, Siegl2025} employ nonlinearities through the use of angle encoding. However, rigorous scaling analyses and quantifiable quantum advantages remain open questions.

A more direct approach encodes nonlinear operations probabilistically within variational quantum algorithms (VQAs), typically amplitude encodinging the fields twice on two separate qubit registers \cite{Lubasch2020, Over2024, Siegl2026}. This framework admits rigorous scaling analysis, has been applied to fluid dynamics \cite{Over2024, Siegl2026,Bengoechea2026} and demonstrated on quantum hardware \cite{Umer2025, Jesus2026-arxiv}. However, the success probability of the nonlinear operation is field-dependent and scales only favorable for specific fields dominated by a high degree of localization \cite{Lubasch2020, Siegl2026} yet not for general nonlinear PDEs \cite{Sarma2024}. 

In this article, we show that naive state-based nonlinear operations pose exponentially decaying success probability for general fields and the time evolution of turbulent fluid flows. To overcome this bottleneck, we develop a hybrid quantum-classical tensor network framework, based on tensor programming  and TT-reconstruction, that offers an alternative implementation of the nonlinear operation.
Benchmarked on turbulent flows, our approach decouples success probability from system size and eliminates the exponential overhead in the number of measurement shots. Crucially, while the scheme adds a classical TT-reconstruction step, it still significantly reduces the computational cost compared to classical tensor network solvers, reaching reductions up to three orders of magnitude. 
Together, these results establish a scalable, resource-efficient framework that brings practical quantum advantage in turbulent CFD within reach.

The rest of this article proceeds as follows: \autoref{sec:method} explains the general method: In \autoref{sec:vqa}, we begin with a concise review of the block encoding with quantum tensor programming  and the variational algorithm originally introduced in \cite{Lubasch2020,Termanova2024, Siegl2026}.
To address the bottleneck of small success probabilities, we introduce an alternative strategy to implement nonlinearities based on tensor programmed diagonal block encodings in \autoref{sec:diagBlock}.
%Next, we discuss the measurement overhead caused by small success probabilities in \autoref{sec:successProbs}. 
The TT-reconstruction required for the diagonal block encodings is detailed in \autoref{sec:TT-reconstruction}. In \autoref{sec:sucProbsMeas} we connect success probability to measurement requirements, and define the practical overhead caused by the operator application.
The introduced method is benchmarked on the example of turbulent fields in \autoref{sec:results}. To this aim, we first describe the governing equations and the considered turbulent flows in \autoref{sec:gov_eq} for which we compute the success probabilities when computing one time step in \autoref{sec:res_succProbs}. To understand potential overheads from the construction of the diagonal block encoding, we analyze and compare the required bond dimensions for the hybrid and the classical tensor network approach in \autoref{sec:res_compression}. This is followed by a praxis test of the TT-reconstruction methods in \autoref{sec:res-TTreconstruct}, and a full scaling analysis and cost comparison with the classical TT implementation in \autoref{sec:comp-tn}.
Finally, we conclude this work and suggest future research directions in \autoref{sec:conclusion}.
\begin{figure*}[bt]
    \centering
    \includegraphics[width=\linewidth]{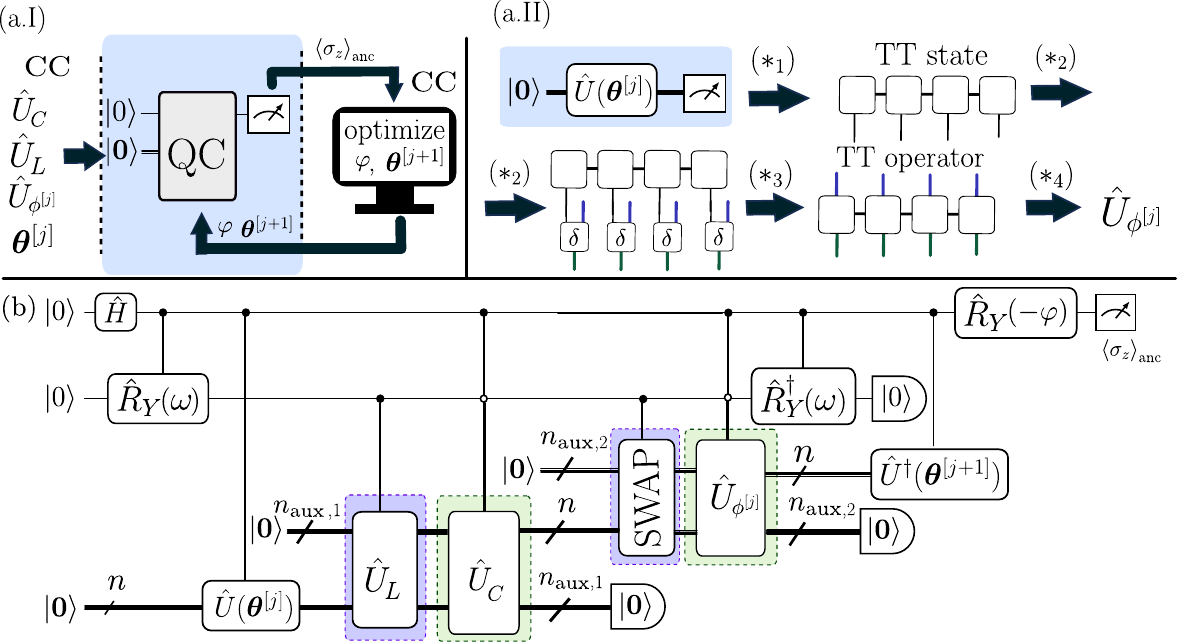}
    \caption{
    (a) Hybrid quantum-classical tensor network algorithm for solving nonlinear PDEs in CFD. The blue box marks operations performed on the quantum computer (QC) while the white areas show quantities stored and operations performed on a classical computer (CC).
    (a.\Romannum{1})  Variational time-stepping loop: the quantum circuit evolves the state using operator unitaries and initial weights from the previous step, while a classical optimizer updates the parameters $\bm{\theta}^{[j+1]}$ and $\varphi$. The cost function $C$ is built from a single expectation value $\expVal$.
    (a.\Romannum{2})  Schematic of the hybrid pipeline to create the diagonal block encoding using the graphical notation of tensor networks \cite{Schollwöck2011}: Starting with the parametrized quantum circuit, $(*_1)$ the state is approximately reconstructed as a tensor train  using one of three possible reconstruction techniques (cf. \autoref{sec:TT-reconstruction}). The reconstructed TT is converted to a diagonal tensor train operator (TTO) by $(*_2)$ applying and $(*_3)$ contracting delta tensors (depicted as $\delta$). Finally, the resulting TTO is $(*_4)$ compiled into unitary gates $U_{\phi^{[j]}}$ via quantum tensor programming\cite{Termanova2024, Siegl2026}.
    (b) Quantum circuit for a single time step of the one-dimensional nonlinear PDE defined in Eq.~\ref{eq:PDE}. Colored blocks denote the linear (violet) and nonlinear (green) operators, combined via a linear combination of unitaries (LCU) using the weighting angle $\omega$ and  the second qubit as auxiliary qubit. Application of SWAP gates is necessary for the linear operation to transport the resulting state on the same register as the nonlinear contribution.
    The first qubit acts as control in an adapted Hadamard test with training parameter $\varphi$; Its expectation value defines the cost function and approaches unity at optimal training. The block encoding unitaries $U_R$ ($R \in {L,C,\phi^{[j]}}$) require auxiliary qubits, where the size of the auxiliary register depends logarithmically on the bond dimension of the TTO and is $n_{\rm aux,1}$ and  $n_{\rm aux,2}$ for the differential operators and the diagonal block encoding, respectively.
    Postselection on the auxiliary register ensures the correct application of the non-unitary evolution. The matrix $\mathds{1}$ in $\hat{O}^{[j]}$ is absorbed in the linear operator block.}
	\label{fig:circuit_scheme}
\end{figure*}
\section{Method} \label{sec:method}
In this section, we present the hybrid quantum-classical tensor network framework for solving nonlinear PDEs in computational fluid dynamics. We begin with a general overview of the considered problem and the algorithmic workflow.
%, followed by detailed descriptions of each component in the subsequent subsections.}

We consider PDEs of the form 
\begin{equation} \label{eq:PDE}
    \frac{\partial \phi}{\partial t} = (\Ol -\hat{O}_{\rm NL})\phi,
\end{equation}
where $\phi=\phi(\bm{x},t)$ is a scalar field solution at the $d$-dimensional spatial position $\bm{x}\in \mathbb{R}^d$ at time $t$ and $\Ol$ and $\hat{O}_{\rm NL}=\hat{O}_{\rm NL}(\phi)$ are linear and nonlinear operators, respectively. For CFD relevant equations, the nonlinear terms typically take the form $(\phi\hat{O}_{C})\phi$, where $\hat{O}_C$ is a time-independent linear differential operator (cf.~\autoref{sec:gov_eq}). We discretize the solution both in space and time such that $\phi^{[j]}=\phi(\bm{x},t=j\Delta t)$ is defined on a grid of size $N^d$ at a specific time step $j$ using a time step size $\Delta t$. Employing explicit Euler time stepping, the operator $\hat{O}^{[j]}=\mathds{1}+\Delta t(\Ol -\Onl)$, evolves the solution by one time step as $\phi^{[j+1]}=\hat{O}^{[j]}\phi^{[j]}$. Here, $\mathds{1}$ is the identity.
The scheme can be generalized to coupled systems of equations and to higher-order time stepping schemes, such as Runge-Kutta, by adapting $\hat{O}^{[j]}$ accordingly. %\nis{higher-order time stepping schemes, such as Runge-Kutta 4 or so} 

To prepare the field $\phi^{[j]}(\bm{x})$ at the $j$-th time step, we  amplitude encode it into an $n=\log_2(N^d)$-qubit register as $\ket{\psi^{[j]}}=1/\theta^{[j]}_0\sum_{l=0}^{N-1} \phi^{[j]}(x_l)\ket{l}$, where $x_l$ are the discrete positions, $\ket{l}$ are the computational basis states
%\Max{Should this not be simply $|l\rangle$?}
and $\theta^{[j]}_0=\|\phi^{[j]}\|_2$ is the $\ell_2$-norm of the field. The state $\ket{\psi^{[j]}}$ is prepared using a parametrized unitary  $U(\bm{\theta}^{[j]})$. %\nis{this big line due to 1/theta looks not so good}
To implement the time evolution via quantum circuits, all differential operators are casted into unitary gates $\hat{O}_R\rightarrow \hat{U}_R$ using quantum tensor programming \cite{Termanova2024, Siegl2026}, where $R$ stands for operator subscripts such as $\rm L$, $\rm C$ or $\rm NL$. In the following we will summarize the linear parts in the time evolution as $\mathds{1}+\Delta t\Ol \rightarrow \hat{U}_{L}$.
The nonlinear multiplication is realized by probabilistically applying a unitary block encoding $\hat{U}_{\phi^{[j]}}$ of the non-unitary diagonal matrix $\hat{D}_{\phi^{[j]}}$, which carries the field values $\phi^{[j]}$ on its diagonal.

The hybrid quantum-classical tensor network scheme is build from two main steps: First, an established variational quantum algorithm \cite{Lubasch2020, Siegl2026} that determines the parameters of the next time step $\bm{\theta}^{[j+1]}$ (see \autoref{fig:circuit_scheme}~(a.I)).
Second, the preparation strategy of the time-dependent unitary $\hat{U}_{\phi^{[j]}}$, which relies on an approximate TT-reconstruction of $\phi^{[j]}$, a subsequent transformation into a TT-operator (TTO) and the compilation into $\hat{U}_{\phi^{[j]}}$ via quantum tensor programming (see \autoref{fig:circuit_scheme}~(a.II)). 

In the following, we first review tensor programmed VQAs
and the previous implementation of nonlinearities. Next, we introduce the alternative treatment of nonlinearities based on tensor-programmed diagonal block encoding, subsequently present three possible techniques to realize the TT-reconstruction step and end with connecting the number of measurements to the success probability.
%% VQA section

\subsection{Tensor Programmed Variational Quantum Algorithms}\label{sec:vqa}
This section provides a short review of the tensor programming-based block encoding and the variational quantum algorithm developed in \cite{Lubasch2020, Termanova2024, Siegl2026}.
We consider a quantum circuit that computes the evolution of one time step as shown in \autoref{fig:circuit_scheme}~(b).
Using linear combinations of unitaries (LCU) \cite{Childs2012} to couple linear and nonlinear terms, the full PDE can be summarized in one circuit. The angles $\omega$ of the LCU are computed as
\begin{equation}\label{eq:LCU}
\omega = 2\arccos(\sqrt{\frac{\beta_1}{\beta_1+\beta_2}}),
\end{equation}
where $\beta_1=\|\mathds{1}+\Ol\|$ and $\beta_2=\|\Onl\|$ are the spectral norms of the operators. This LCU coupling naturally extends to additional terms, necessary e.g. in coupled PDEs.

In order to apply the non-unitary operators $\hat{O}_R/c$ to the quantum state $\ket{\Psi^{[j]}}$, we employ quantum tensor programming described in detail in \cite{Termanova2024, Siegl2026}. Here, the subnormalization constant $c\geq\|\hat{O}_R\|$ and $\|\hat{O}_R\|$ is the spectral norm of the target operator.
Quantum tensor programming is an efficient strategy to realize block encodings \cite{Gilyen2019} by approximately translating a TTO into unitary gates. The required circuit depth scales polynomially with the bond dimension of the target TTO %and linearly with the qubit number 
\cite{Termanova2024, Siegl2026} and TTOs provide an exact low-rank representation for many differential operators, with a bond dimension independent of the system size \cite{Kazeev2012, Oseledets2010, Kiffner2023}.
The computed unitaries $\hat{U}_R$
are applied on the state register and an auxiliary register of size $n_{\rm{aux}}$ (see \autoref{fig:circuit_scheme}~(b)), where $n_{\rm{aux}}$ scales logarithmically with the bond dimension of the unitary TTO \cite{Termanova2024, Siegl2026}.
After the operator application, the auxillary register is measured and postselected on $\ket{\bm{0}}_{\rm{aux}}$. Only when measuring $\ket{\bm{0}}_{\rm{aux}}$, the application of the target operator was successful and the solution was correctly evolved by one time step. Otherwise the circuit evaluation needs to be discarded.
%For many differential operators, TTOs provide an exact low-rank representation \cite{Kazeev2012, Oseledets2010, Kiffner2023}, with a bond dimension independent of the system size wich translates into a low-depth quantum circuit \cite{Termanova2024, Siegl2026}. 
%To enable block encoding, we need to devide $\hat{O}_R$ by a
%subnormalization constant i.e. $\hat{O}_R/c$
The success probability of the operator application is 
\begin{equation}\label{eq:succ_probs}
\succProbs=\|\hat{O}_R\ket{\Psi^{[j]}}\|_2^2/c^2,
\end{equation}
being maximal for  $c =\beta_{\rm opt}=\|\hat{O}_R\|$.
%An efficient strategy to realize block encodings for differential equations is quantum tensor programming described in detail in \cite{Termanova2024, Siegl2026}, which approximately translates a TTO into unitary gates. 
%Specifically, we utilize the formulation described in detail in \cite{Termanova2024, Siegl} to block-encode non-unitary operators as a unitary-and-projection operation for relevant TT operators from the linear terms, as well as a diagonal TT operator that encodes the TT state in its diagonal that are used in this article. 

To determine the parametrization of the next time step $\bm{\theta}^{[j+1]}$, we minimize the cost function $\mathcal{C}=-\bra{0}\hat{U}^{\dagger}(\boldsymbol{\theta}^{[j+1]})\hat{O}^{[j]}\ket{\psi^{[j]}}$ (see \autoref{fig:circuit_scheme}~(a.I))
The parameters of the next time step $\bm{\theta}^{[j+1]}$ are initialized with $\bm{\theta}^{[j]}$, to avoid the occurrence of barren plateaus and improve trainability \cite{Siegl2026}.
% specifically the variational time stepping rotine, block encoding and the concept of quantum tensor programming }
%a variational quantum algorithm used to compute the next time step (I) and a quantum tensor programming routine (II) that is necessary to compute the non-linearity without exponentially increasing measurement requirements.
%To prepare the field $\phi(\bm{x},j)$ at the $j-$th time step, we  amplitude encode it into an $n=\log_2(N^d)$-qubit register as $\ket{\psi^{[j]}}=\frac{1}{\theta^{[j]}^0}\sum_{l=0}^N \phi(x_l,j)\ket{x_l}$, where $\ket{x_l}$ are the computational basis states and $\theta^{[j]}^0=\|\phi^{[j]}\|$ is the norm of the classical field. The state $\ket{\psi^{[j]}}$ is prepared using a parametrized unitary  $U(\bm{\theta}^{[j]})$.

The cost function $\mathcal{C}$ can be evaluated by measuring an expectation value $\expVal$ from a Hadamard test adapted for probabilistic operator encoding \cite{Siegl2026}. For the adapted Hadamard test, we substitute the second Hadamard gate with $R_Y(\varphi)$ , as shown in \autoref{fig:circuit_scheme}~(b). The optimal choice of $\varphi$ is given by $\phiOpt=2\arctan(\sqrt{\alpha_{\rm{succ}}})$ \cite{Siegl2026}.
Using $\phiOpt$, the cost function will reach $\mathcal{C}=-1$ in case of optimal training of $\bm{\theta}^{[j+1]}$. Furthermore, the fidelity $\mathcal{F}$ of the optimal solution and the actually trained solution is given by 
\begin{equation}
\mathcal{F} = \frac{(\alpha_{\text{succ}}(\expVal+\cos(\phiOpt))+\expVal-\cos(\phiOpt))^2}{4\alpha_{\text{succ}}\sin^2(\phiOpt)},
\end{equation}
providing an efficient measure of convergence without the need of comparison to classical data \cite{Siegl2026}. The parameter $\varphi$ is optimized itself, and used to compute $\succProbs$.
The normalization of the next time step is given in terms of the success probability as $\theta^{[j+1]}_0=\beta_{\rm{full}}f_{\hat{O}^{[j]}}\theta^{[j]}_0$ where $\beta_{\rm{full}}=\beta_1+\beta_2$ (cf. \autoref{eq:LCU})
%\nis{Whats beta full? introduced?} \pia{like this?}
and $f_{\hat{O}^{[j]}}$ is a norm correction factor that is related to $\succProbs$ and $\varphi$ via \cite{Siegl2026}
\begin{equation}
f_{\hat{O}^{[j]}}=\frac{1+\alpha_{\text{succ}}}{2\sin(\varphi)-(\sqrt{\alpha_{\text{succ}}}-\frac{1}{\sqrt{\alpha_{\text{succ}}}})\cos(\varphi)}.
\end{equation}

%\subsection{Success Probabilities and its Measurement Overhead} \label{sec:successProbs}
%\subsection{Success Probability and Measurement Requirements}\label{sec:succProbs}
Crucially, the classical-quantum feedback loop prevents exponential decay of $\succProbs$ over the number of time steps, unlike in purely quantum time-stepping schemes \cite{Mezzacapo2015, Zhao2023, Bharadwaj2025}.
Furthermore, for the block encoding of standard linear differential operators, the average success probability over Haar-random and uniformly random states remains constant with system size \cite{Termanova2024, Siegl2026}.

Prior quantum implementations \cite{Lubasch2020, Over2024, Siegl2026} compute the nonlinearity by implementing $\ket{\psi^{[j]}}$ twice  using the same parametrized circuits on two distinct quantum registers, followed by a ladder of CNOT gates and a mid-circuit measurement of the second register. While exact, it only scales favorably  for highly localized states dominated by narrow peaks \cite{Lubasch2020, Siegl2026}. Instead, for general fields it leads to an exponential decrease in the success probability as shown for various PDEs in \cite{Sarma2024} and demonstrated here for CFD applications (see \autoref{sec:res_succProbs}).

%The angle $\omega$ can be computed directly from the used $\beta$ of the operators
%\begin{equation}
%\omega = 2\arccos(\sqrt{\frac{\beta_1}{\beta_1+\beta_2}}),
%\end{equation}
%where $\beta_1=\|\mathds{1}+\hat{O}^l\|$ and $\beta_2=\|O^{nl}(\phi)\|$.
%For higher dimensional use cases additional auxillary qubits are required. The computation of angles follows the standard LCU approach.

%\textcolor{blue}{potentially shift to later in the text:}
%To combine all contributions of the partial differential equation in one quantum circuit, we make use of a linear combination of unitaries (LCU). For \autoref{eq:NSE_component}, the resulting composed circuit is depicted in \autoref{fig:circuit_scheme}~(b).

%and is given for $2$ and $3$ dimensional equations in Appendix -----. 

\subsection{Diagonal Block Encoding for Nonlinear Terms}\label{sec:diagBlock}
We circumvent the bottleneck of exponentially decaying success probabilities by reformulating the nonlinear multiplication as the application of a diagonal matrix  $\hat{D}_{\phi^{[j]}}$ to the quantum state (see \autoref{fig:circuit_scheme}~(b)). 
Implementing the diagonal operators $\hat{D}_{\phi^{[j]}}$ using quantum tensor programming, requires knowledge of $\phi^{[j]}$ beyond the variational parameters $\bm{\theta}^{[j]}$. 
We address this by reconstructing an approximate TT-state representation of $\phi^{[j]}$, converting it into a TTO that represents $D_{\phi^{[j]}}$ and compiling it via quantum tensor programming into unitary gates $U_{\phi^{[j]}}$ (\autoref{fig:circuit_scheme}~(a.II.)). Whenever there is a TT-state representation, it can be transformed into a TTO that encodes a diagonal matrix, using a contraction of each TT-state tensor with a delta tensor \cite{Gourianov2022}. 

The normalization constant for the block encoding is chosen as $c_{\rm diag}=\beta_{\rm opt}=\|\hat{D}_{\phi^{[j]}}\|=\max|\phi^{[j]}|$ which depends only on the physical problem and is independent of the grid size. In practice, $c_{\rm diag}$ can be approximately sampled from the classically reconstructed TT \cite{Xiao2026-arxiv} or adaptively tuned during tensor programming compilation as described in \cite{Termanova2024, Siegl2026}.
In contrast, the previous state-based implementation of the nonlinearity implements a block encoding of $D_{\phi^{[j]}}$ using a map from $\ket{0}\rightarrow \ket{\psi^{[j]}}$ on a second qubit register. This leads to a native subnormalization constant of $c_{\rm{vec}}=\|\phi^{[j]}\|_2 \sim \sqrt{N^d}$, which is the $\ell_2$-norm of the classical field and hence typically scales with system size.
Consequently, when computing the square of a uniform one-dimensional field 
$\phi(x)=1$, the diagonal block encoding yields 
$\succProbs=1$, whereas the state-base Hadamard product gives $\succProbs=1/N$.
This scaling advantage does not depend on the complexity of the field and is equally true for $\phi$ composed of uniform random distributed real numbers between $0$ and $1$ or $\phi$ being drawn from a real Haar random distribution.
In the large $N$ limit, for the uniform random (real Haar random) case, the success probability asymptotically approaches $\succProbs = 3/5$ ($\succProbs\propto 1/\log_2(N)$) for the diagonal block encoding and $\succProbs= 9/(5N)$ ($\succProbs\propto1/N$) for the state-based Hadamard product (see Appendix~\ref{ap:randhadamardprod} for the derivations).
Furthermore, the advantage extends to the time evolution of turbulent fields, as demonstrated in \autoref{sec:results}.
While it is possible to construct diagonal block encodings directly from quantum states \cite{Rattew2023-arxiv}, these block encodings naturally have $c=c_{\rm{vec}}$ and hence can not stabilize  $\succProbs$.

Once $U_{\phi^{[j]}}$ is prepared, the quantum circuit executes the complete time step—including the nonlinear multiplication—without overhead caused by contractions and truncations which dominate the cost in classical tensor network solvers.
\subsection{Tensor Train Reconstruction} \label{sec:TT-reconstruction}
In order to construct the diagonal operator $\hat{D}_{\phi^{[j]}}$, we consider three distinct strategies to obtain a TT-reconstruction  of $\phi^{[j]}$ from the parametrized quantum circuit $\hat{U}(\bm{\theta}^{[j]})$: 
The first, tensor cross interpolation (TCI), constructs a low-rank tensor approximation by querying state amplitudes on a sparse, cross-shaped subset of grid points \cite{TCIPaper}. 
The amplitude queries required for TCI can be performed using either Hadamard tests \cite{Lubasch2020} or amplitude estimation \cite{Brassard2000QuantumAA}.
Second, efficient quantum state tomography (EST) as introduced in \cite{Cramer2010} avoids the exponential overhead of full state reconstruction by measuring reduced density matrices on sequential qubit subsets. Each measured subset is disentangled from the remaining system via a unitary transformation, and the accumulated unitaries are used to assemble the target TT. The number of qubits measured per step scales logarithmically with the target bond dimension.
The third approach, TT state simulation (TTS), classically simulates the parameterized circuit $\hat{U}(\bm{\theta}^{[j]})$ using tensor trains.  Crucially, the classical TTS only performs the TT simulation of the parametrized circuit of the already computed time step but does not perform the computation of the full evolution step. This leads to a significantly different computational cost of the hybrid approach using TTS compared to a classical TT simulation of the time evolution (cf. \autoref{sec:comp-tn}).

The computational cost of state reconstruction is fundamentally tied to the possible compression of 
the target state, as it directly depends on the bond dimension. Hence, whenever $\phi^{[j]}$ admits an efficient TT representation with low bond dimension, both state extraction and subsequent block encoding remain efficient. The cost dependence on the bond dimension rather than on system size makes the classical reconstruction step not negate the quantum advantage for the time evolution of turbulent flows (cf. \autoref{sec:comp-tn}).%, while keeping its cost significantly below that of a classical TT computation of the time evolution.

Implementation details of all three methods are provided in Appendix~\ref{app:tt-reconstruction}. Their resource trade-offs and performance is explained and benchmarked on turbulent flow fields in Sec.~\ref{sec:res-TTreconstruct},
showing a significantly reduces classical cost when using TT-reconstruction, compared to a purely classical TT time evolution routine (cf. \autoref{sec:comp-tn}). While TTS increases classical computational overhead compared to TCI and EST, it entirely eliminates the need for quantum measurement routines. 
\subsection{Success Probability and its Measurement Overhead}\label{sec:sucProbsMeas}
The success probability $\succProbs$ directly governs the measurement overhead of the hybrid algorithm. In the adapted Hadamard test, the expectation value $\expVal$ approaches unity for optimally trained parameters. However, the accurate estimation of $\phiOpt$, and consequently $f_{\hat{O}^{[j]}}$ requires an increasingly accurate resolution of $\expVal$ as $\succProbs$ decreases (see Appendix \ref{ap:error-prop} for details). This leads to a dependence of the required measurement shots $m$ on the success probability as 
\begin{equation}
    m \propto 1/\succProbs.
\end{equation}

Mid-circuit measurements and postselection are not strictly required when block encodings are embedded in the Hadamard test, as the Hadamard test inherently filters the $\ket{\bm{0}}_{\rm{aux}}$ subspace \cite{Lubasch2020}. However, this coherent filtering does not reduce the measurement overhead: it suppresses the maximal observable signal to $\expVal=\sqrt{\succProbs}$, preserving the same shot scaling $ m \propto 1/\succProbs$  \cite{Lubasch2020}.  We therefore presented the method using explicit post-selection for clarity, noting that the optimal choice between mid-circuit measurements and coherent filtering in practice will depend on hardware-specific constraints and noise resilience.

Given the block-encoding based implementation of the nonlinearity and its stabilization of the success probability, the number of required measurements solely depends on the target accuracy of the result (cf. \autoref{sec:comp-tn}) but becomes independent of system size.
\section{Application to Turbulent Flows}\label{sec:results}
To understand the potential of the introduced method, we benchmark it at the relevant example of the time evolution of turbulent flows. We will show stable success probabilities of $15-30\%$ across increasing Reynolds numbers and system sizes when employing the tensor-based diagonal block encoding. Furthermore, we show that the computational cost of the classical TT-reconstruction step remains several orders of magnitude below the cost of a fully classical tensor network routine.
\subsection{Governing Equations and Use Cases}\label{sec:gov_eq}
To this aim, we consider the set of coupled differential equations in $d$ spatial dimensions
\begin{equation} \label{eq:NSE_component} 
	\frac{\partial u_i}{\partial t} = \nu \sum_{k=1}^d\frac{\partial^2 u_i}{\partial x_k^2} - \sum_{k=1}^d u_k \frac{\partial u_i}{\partial x_k}, 
\end{equation}
which corresponds to \autoref{eq:PDE} when setting $\phi=u_i$, $\Ol= \nu \sum_{k=1}^d\frac{\partial^2}{\partial x_k^2}$ and $\hat{O}_{\rm NL}(\mathbf{u})=\sum_{k=1}^d u_k \frac{\partial}{\partial x_k}$.
Here $u_i$ denotes the $i$-th component of the $d$-dimensional velocity vector $\bm{u}$, the derivatives are taken along the $k$-th direction of the $d$-dimensional space, and $\nu$ is the viscosity.
Equation~\ref{eq:NSE_component} corresponds to the Burgers' equation and the momentum equation in the incompressible Navier-Stokes equation, when the well-established fractional step method is employed \cite{Chorin1968, vanHuelst2026}.
%\nis{well-established sounds over-selling here, probably should cite Chorin's original paper here and not vanHuelst (in case its cited anyways somewhere else already?);}

We evaluate the algorithm on two canonical turbulent benchmarks: the two-dimensional temporally developing jet (TDJ) and the three-dimensional Taylor--Green vortex (TGV). Both flows have been extensively studied in classical tensor network literature \cite{Gourianov2022}.
Reference fields are generated via direct numerical simulation at multiple Reynolds numbers ($\rm Re$) and grid resolutions using the DNS Code provided in \cite{Gourianov2022Code}. These DNS solutions serve as ground truths for benchmarking the success probability, bond dimension requirements, and reconstruction accuracy of the hybrid scheme. Details on the simulation parameters and the DNS code are provided in Appendix~\ref{ap:turbDat}.

\subsection{Success Probability for Turbulent Flow Evolution}\label{sec:res_succProbs}
Here, we show the computed success probabilities and compare the previous state-based implementation of the nonlinearity with the alternative block-diagonal encoding. To this aim we consider the success probability when computing the time evolution of a turbulent flow.
By combining LCU with quantum tensor programming, the full time-evolution operator for a single step can be compiled into a single quantum circuit (cf. \autoref{sec:vqa}). This yields a composite success probability that directly governs the measurement overhead.
For the momentum equation (\autoref{eq:NSE_component}) and explicit Euler time stepping, the success probability for evolving the 
$i$-th velocity component is
\begin{equation}
\begin{aligned}
    &\succProbs(u_i|\Delta t,\nu,\bm{u}) = \\
    &\left\|\frac
{1}{\beta_{\rm full}} \left(\left(\mathds{1}+\sum_{k=1}^d\frac{\Delta t}{\nu} \frac{\partial^2}{\partial x_k^2}\right) u^S_i - \Delta t\left( \sum_{k=1}^d\tilde{u}_k\frac{\partial u^S_i}{\partial x_k}\right )\right)\right\|^2,
\end{aligned}
\end{equation}
with 
\begin{equation*}
\beta_{\rm full} = \beta_L + \sum_{k=1}^d\beta_{u_k} \beta_{\partial_{x_k}}.
\end{equation*}
Here, $\tilde{u}_k$ can either be $u^{S}=u/\|u\|_2$ that corresponds to the
$\ell_2$-normalized state used in prior state-based implementations, or $u^D=u/\max(|u|)$ where the field is divided by its maximum absolute value, as required for the diagonal block encoding. The subnormalization constants $\beta_L$, $\beta_{u_k}$ and $\beta_{\partial_{x_k}}$ are the spectral norms of the linear operator contributions, the diagonal field encodings and the spatial derivatives, respectively, leading to optimal success probabilities.
%Optimal success probabilities are achieved when each $\beta_{M}$ equals the spectral norm of the corresponding operator  $M$ \cite{Termanova2024}.
\begin{figure}[bt]
    \centering
    \includegraphics[width=\linewidth]{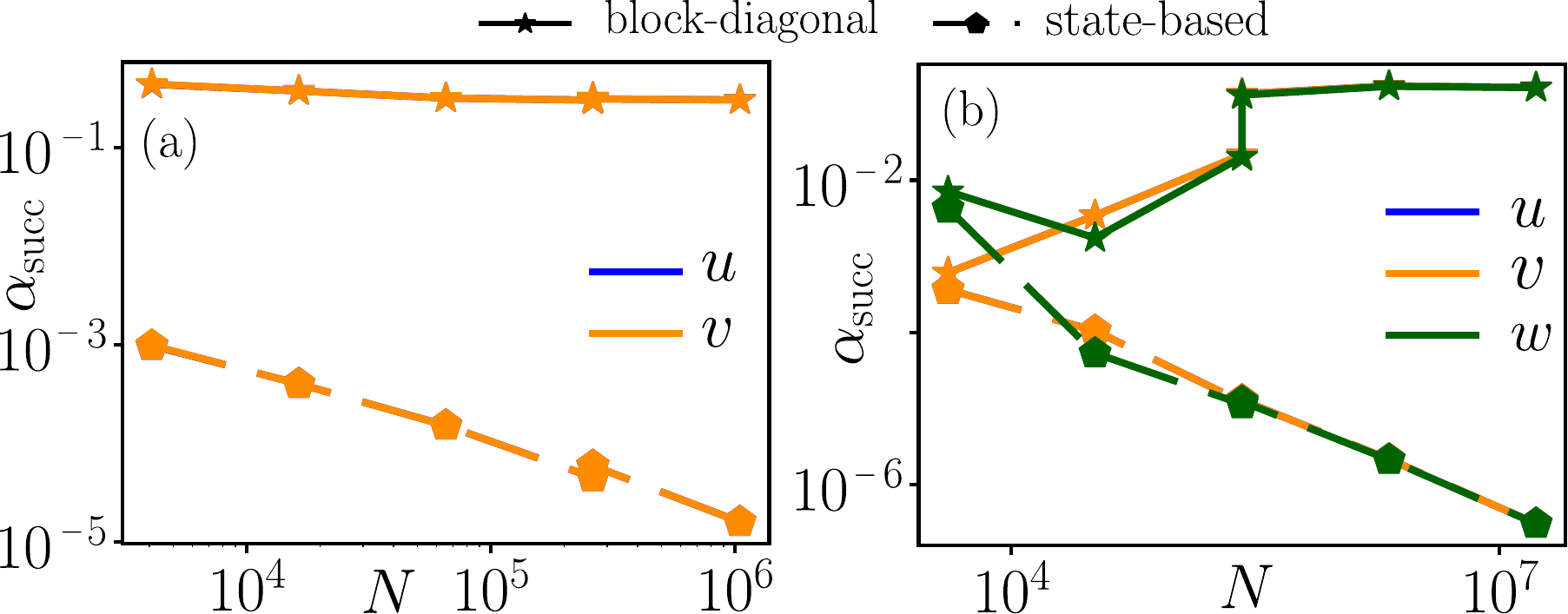}
    \caption{
     Optimal success probabilities $\succProbs$ over the number of grid points $N$. Solid lines with stars denote the diagonal block encoding approach, while dashed lines with pentagons represent the state-based Hadamard product
     %(a) Success probability of the term $\phi^2$, where $\phi$ is a random vector drawn from a uniform distribution (olive) or a constant vector of ones (purple). 
     (a) Success probability for a single  Euler time step of the momentum equation for the two velocity components $u,~v$  of the two-dimensional TDJ flow and (b) the three velocity components $u,~v,~w$ of the three-dimensional TGV flow. In panels (a) and (b), increasing grid resolution corresponds to higher Reynolds numbers. Details on the simulation parameters are provided in Appendix \ref{ap:turbDat}. The success probabilities of the velocity components $u$ and $v$ are very similar and ly visibly on top of each other across all $N$.}
\label{fig:alpha_opt}
\end{figure}

Figure~\ref{fig:alpha_opt} shows the success probabilities for one Euler time step applied to the two-dimensional TDJ and three-dimensional TGV across increasing $\rm Re$. The state-based approach exhibits a strong 
%$1/N$ \nis{$N$ or $N^d$?} 
decay with the number of grid points, whereas the diagonal block encoding maintains a stable or even increasing success probability. For the TGV, we observe lower success probabilities for low $\rm Re$ using the block-diagonal encoding. At these low $\rm Re$, turbulence does not yet emerge and the fields are dominated by large-scale structures.
The results show, that the hybrid routine retains measurement efficiency for large-scale simulations at high Reynolds numbers. 

\subsection{Compression Rates}\label{sec:res_compression}
In addition to success probability, the bond dimension required for the TT-reconstruction is a key factor determining the computational cost of the hybrid scheme.
For the momentum equation (\autoref{eq:NSE_component}), the fields mapped to the diagonal block encoding 
$\hat{D}_{u_k}$ %\nis{maybe hat on D?} 
appear only as prefactors in the nonlinear advection term: 
\begin{equation} \label{eq:adv-term}
A_i = u_k \frac{\partial u_i}{\partial x_k} \rightarrow \hat{D}_{u_k} \frac{\partial u_i}{\partial x_k}. 
\end{equation}
Consequently, only $\hat{D}_{u_k}$ is affected by truncation errors introduced in the TT-reconstruction, while the accuracy of  $u_i$ solely depends on the quantum circuit implementation of the field.
In contrast, classical tensor network solvers must truncate both fields and apply additional truncation steps after each operator application to control bond dimension growth \cite{Schollwöck2011}. 

\begin{figure}[bt]
    \centering
\includegraphics[width=\linewidth]{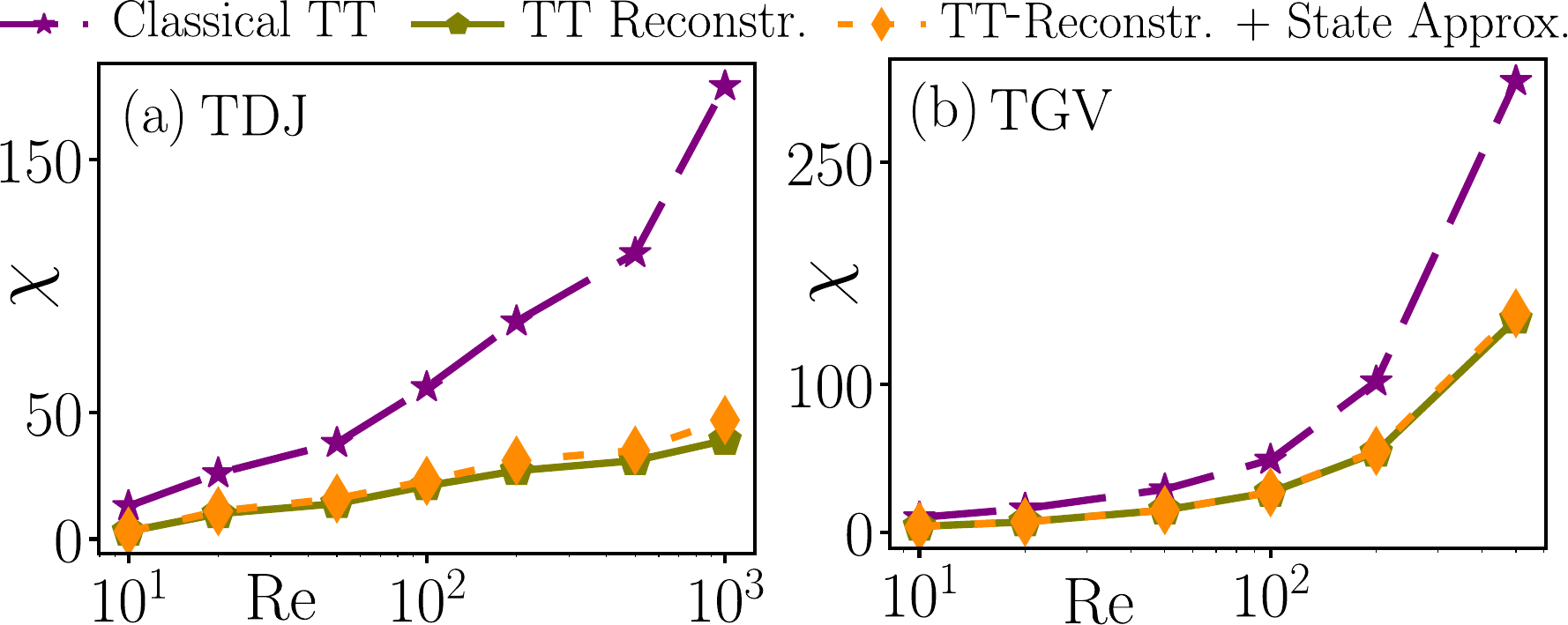}
    \caption{
     Required bond dimensions for evaluating all nonlinear advection terms $A_i$ within a target $\ell_2$-error $\epsilon_{\ell_2}\leq0.01$.
     We consider three cases: 
     First, \textit{Classical TT} (deterministic evaluation of the complete nonlinear term with TTs using bond dimension $\chi_{\rm TN}$), second,
     \textit{TT-Reconstruction}(truncation of the prefactor field $u_k$ to $\chi_{\rm{base}}$, reflecting the classical reconstruction step in the hybrid approach), 
     and  third, \textit{TT-Reconstruction + Circuit Approx.} (truncation of the prefactor field $u_k$ to $\chi_{\rm{base}}$, additionally accounts for finite quantum circuit accuracy by approximating the differentiated field $u_i$ in $A_i$ as a TT with 
$\chi_{\rm TN}+0.1\chi_{\rm TN}$). The slightly increased bond dimension of the differentiated field compared to $\chi_{\rm TN}$ accounts for the higher representation capability of the quantum circuit \cite{Siegl2026}.
Panels (a) and (b) show the bond dimension requirements for the two-dimensional TDJ and three-dimensional TGV flows, respectively, across increasing Reynolds numbers $\rm Re$. }
     \label{fig:chi_min}
\end{figure}
To estimate the error between a term $K_i$ and its approximation $\tilde{K_i}$ we consider the relative $\ell_2$-error 
\begin{equation} \label{eq:l2-error}
    \epsilon_{\ell_2}=\frac{\|K_i-\tilde{K_i}\|_2}{\|K_i\|_2}.
\end{equation}
In the following, $K_i$ will refer to the field components $u_i$, their derivative or the nonlinear advection terms $A_i$.
In general we observe, that representing a field $\phi$ approximately up to a target $\epsilon_{\ell_2}$
requires a lower bond dimension than representing its gradient $\partial \phi/\partial x$ up to the same accuracy  (cf. Appendix~\ref{app:bond-dim-scaling}).
We have observed truncation-induced oscillations in the field which are small in magnitude but are presumably the reason for larger deviations in the derivative. 
%\nis{i think it needs rather an interpretation, like interprete it this way, but not proven}

We denote $\chi_{\rm{base}}$ as the minimal bond dimension required to represent  $\hat{D}_{u_k}$ such that all $A_i$ are computed within a target $\epsilon_{\ell_2}$ and $\chi_{\rm TN}$ 
%\nis{why c? classical? what you mean by full afterwards? untruncated?} \pia{yes}
as the bond dimension required for the classical TT evaluation of all $A_i$ up to the same accuracy. As shown in \autoref{fig:chi_min}, $\chi_{\rm{TN}}$ grows significantly faster than $\chi_{\rm{base}}$ with increasing Reynolds number. For the three-dimensional TGV, $\chi_{\rm{base}}$ remains about a factor $\approx2$ smaller than $\chi_{\rm{TN}}$ across all resolutions. For the two-dimensional TDJ, this saving increases to a factor of $\approx3$.
%This compression gap persists throughout time evolution (Appendix~\ref{app:bond-dim-scaling}).
Importantly, the required $\chi_{\rm{base}}$ increases only slightly, when accounting for approximation errors in the amplitude-encoded field itself, preserving the favorable ratio (see \autoref{fig:chi_min}). 
  %For the turbulent flows considered, we compute the bond dimension required to represent the nonlinear terms $A_i$ with $\epsilon_{\ell_2}\leq0.01$
%When considering the error per time step, the target accuracy $\epsilon_{\ell_2}$ for the nonlinear term can be scaled with $\Delta t$, reflecting its linear weighting in \autoref{eq:NSE_component}. Nevertheless, maintaining a small error in the nonlinear term is essential to prevent unphysical error accumulation and numerical instability (cf. Appendix~\ref{app:bond-dim-scaling}).
\subsection{Tensor Train Reconstruction of Turbulent Fields} \label{sec:res-TTreconstruct}
\begin{figure}[bt]
    \centering
    \includegraphics[width=\linewidth]{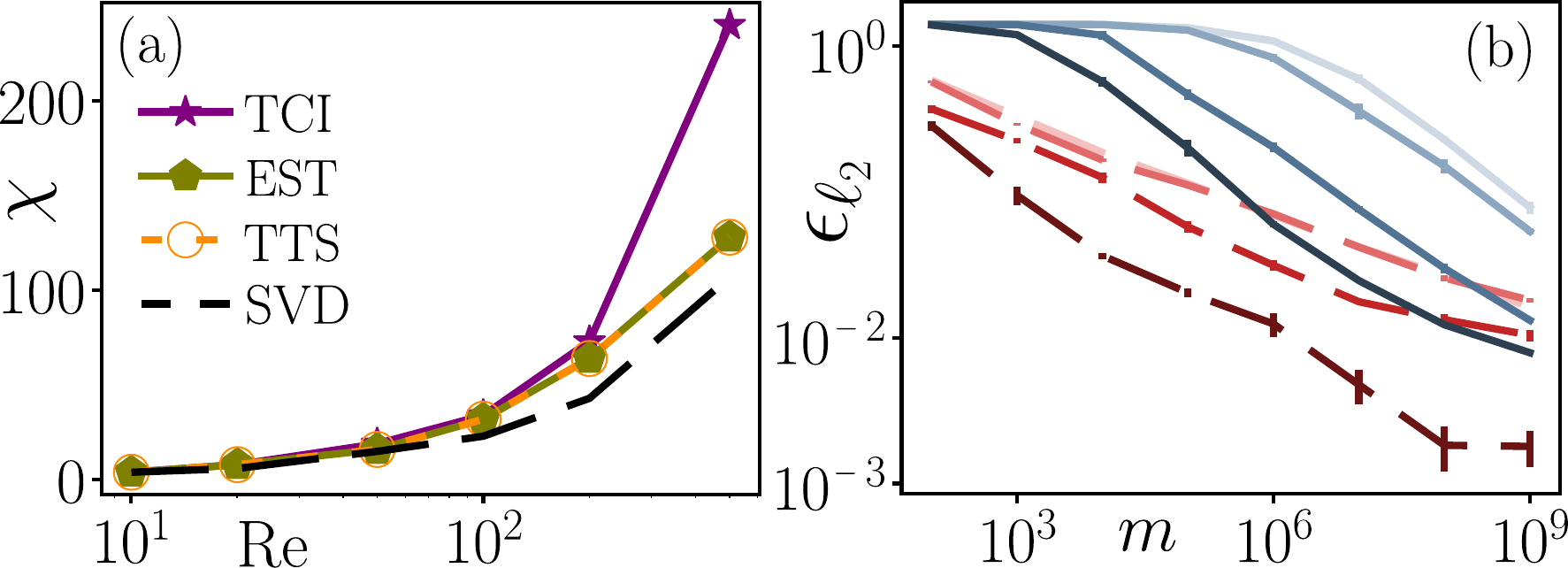}
    \caption{Evaluation of the TT-reconstruction techniques for the three-dimensional Taylor--Green vortex. (a) Required bond dimension $\chi$ to represent the first component of the velocity vector $u$ up to the relative $\ell_2$-error $\epsilon_{\ell_2} \leq 0.01$ across increasing Reynolds numbers $\rm Re$. For TCI, $\chi$ is chosen such that the mean error over 100 independent runs satisfies $\epsilon_{\ell_2} \leq 0.01$. (b) $\epsilon_{\ell_2}$ over $10$ independent reconstructions as a function of measurement shots $m$ per expectation value, emulating finite-shot sampling noise. Line shades indicate increasing Reynolds number ($\rm{Re}={10,20,50,100}$, dark to light) using the  EST (red, dashed lines) and the TCI (blue solid lines). Error bars represent the standard deviation. We use the bond dimension depicted in panel (a), hence, $\epsilon_{\ell_2}$ reachable with finite shot number is lower bounded by the accuracy reachable with the state-vector simulation.
    Simulation parameters and implementation details are provided in Appendix~\ref{app:tt-reconstruction}.}
	\label{fig:mps-tomo}
\end{figure}
The TT-reconstruction is a critical component of the hybrid pipeline. Here, we evaluate the three TT-reconstruction strategies introduced in \autoref{sec:TT-reconstruction}, comparing their bond dimension requirements and measurement overhead for turbulent flow fields. 
Figure~\ref{fig:mps-tomo}~(a) depicts the required bond dimension for TCI, EST and TTS of the parametrized circuit. The SVD-based TT decomposition serves as the theoretical baseline for optimal compression with bond dimension $\chi_{\rm{base}}$.
TCI exhibits a pronounced increase in required bond dimension $\chi_{\rm TCI}$ as $\rm{Re}$ increases, demanding more than twice the baseline for large $\rm{Re}$.
In contrast, we see that both EST and TTS remain close to the optimal baseline, requiring only a power-of-two ceiling $\chi_{\rm{ceil}}=2^{\lceil\log_2\chi_{\rm{base}}\rceil}$.
%Details on the choosen parameters for TCI, the circuit simulated for the mps-simulation are provided in \autoref{ap:}.

Both TCI and EST are expected to require the computation of $\mathcal{O}(\chi_{\rm{ceil}}^2)$ expectation values, each sampled with a finite number of measurement shots. 
Figure~\ref{fig:mps-tomo}~(b) compares the reconstruction accuracy under finite-shot sampling, where each expectation value is estimated with a limited number of measurements.
For TCI, the error exhibits a broad plateau at high $\rm Re$. This behavior stems from amplitude encoding normalization: as grid resolution increases, individual state amplitudes scale as 
$\mathcal{O}(1/\sqrt{N^d})$, degrading the signal-to-noise ratio and increasing shot requirements. Once sufficient shots are allocated to overcome this noise floor, the error decays as $\propto 1/\sqrt{m}$ until saturating to its lowest possible error, which is determined by the provided bond dimension. In contrast, EST avoids this initial plateau by measuring reduced density matrices on qubit subsets rather than individual amplitudes. %Consequently, its shot scaling depends on the field's entanglement complexity (reflected in $\chi_{\rm{base}}$) rather than the grid size $N^d$. 
While the error in each individual density matrix scales as $\approx1/\sqrt{m}$, the accumulated reconstruction error decreases more slowly due to error accumulation across sequential disentangling steps and also saturates at the lower error bound set by the bond dimension. Advanced sampling protocols, such as classical shadows \cite{Qin2025-arxiv} or more involved postprocessing \cite{Cramer2010} could further mitigate these measurement requirements in future implementations.
For both approaches we have considered a sampling-based estimation of the expectation values (cf.~Appendix~\ref{app:tt-reconstruction} for details). 

Comparing EST and TCI, the EST shows an overall better performance, requiring lower bond dimensions and a reduced number of measurement shots per expectation value.
%\pia{As we have used the required bond dimension to reach $\epsilon_{\ell_2}\leq 0.01$, increasing the number of shots will not allow for a deacy below the original error. As ETS always uses the ceiled bond dimension, its final $\epsilon_{\ell_2}$ is usually lower than that of TCI. Overall, ETS seems to be superiour for obtaining the TT-reconstruction, as it required a lower bond dimension and a lower number of measurement shots per projection.}
%The required number of measurements is expected to be reduced when employing more sophisticated routines as amplitude estimation or more complex postprocessing \cite{Cramer2010} and will be addressed in  future work.
However, given the substantial shot requirements for measurement-based reconstruction, TTS emerges as a highly viable alternative. Its classical computational cost scales as $\chi_{\rm{ceil}}^3$, matching the quantum tensor programming compilation step and introducing no additional asymptotic overhead.

Future optimizations in diagonal block encoding may further reduce compilation costs, potentially shifting the optimal trade-off toward measurement-based reconstruction. Nevertheless, TTS provides a robust, measurement-free pathway to the efficient TT-reconstruction used for the preparation of the nonlinear operators. 
%\nis{too long, wording, and many commas in this sentencen. split apart}

\subsection{Scaling Analysis and Comparison to Classical Tensor Network Methods} \label{sec:comp-tn} To quantify the resource efficiency of the hybrid framework, we decompose the total computational cost $T_{\rm H}$ per time step into quantum and classical contributions: 
\begin{equation}
    T_{\rm H} = T_{\rm{QC}} + T_{\rm CL}
\end{equation}
The quantum cost $T_{\rm{QC}}$ comprises circuit execution and measurement overhead. For a circuit of depth $d_{\rm{circuit}}$, the circuit execution cost scales as $\mathcal{O}(d_{\rm{circuit}})$. 
Combined with the shot requirement $m\propto1/(\epsilon_{\rm meas}^2\succProbs)$ to estimate all relevant quantities to precision $\epsilon_{\rm meas}$ (cf. Appendix~\ref{ap:error-prop} for details) the total quantum overhead scales as
%The cost of the quantum circuit evaluations $T_{qc}$ is composed by the cost of the circuit itself $O(n_q)\cdot d_{circuit}$ times the number of required measurements depending on $\succProbs$ and the target accuracy $\epsilon$ 
%of the fidelity measure and norm correction being 
%$O(1/\epsilon^{(2)}\cdot\succProbs)$.
\begin{equation} 
T_{\rm{QC}} \sim \left( \frac{ d_{\rm circuit}}{\epsilon_{\rm meas}^2 \alpha_{\rm succ}} \right). \end{equation}
For the diagonal block encoding $\succProbs=\mathcal{O}(1)$ and $d_{\rm circuit}$ is upper bounded by $\chi^2$ \cite{Lubasch2020, Gourianov2022}, where $\chi$ is the bond dimension of a TT with the same target accuracy.
In practice, the circuit depth can be reduced below $\chi^2$ as shown for quantum states \cite{Haghshenas2022} and turbulent fields \cite{Siegl2026}.
More sophisticated circuit designs and training strategies are expected to further decrease the depth of the quantum circuits, potentially reaching the predicted target depth of $\rm{poly}(n)$ \cite{Hashizumeedgeofchaos2026,Meng2026}.

%For the classical procedure, the required resources are composed from the two dominant steps. First, the TT reconstruction, where we assume the cost of the TT-circuit simulation scheme in the following. Second, the quantum tensor programming routine to create the unitary gate set that encodes the block encoding of $D_\phi$, which scales as $\chi_{\rm{min}}^3$ \textcolor{blue}{actually the target TT-operator bond dimension $Z^3$}. The depth can be reduced below $\chi^2$ as shown for quantum states \cite{Haghshenas2022} and turbulent fields \cite{Siegl2026}.
%More sophisticated circuit designs and training strategies are expected to further decrease the depth of the quantum circuits, potentially reaching the predicted target depth of $\rm{poly}(n)$ \cite{Meng2026}.

The classical cost $T_{\rm{CL}}$ is dominated by two steps: (i) TT simulation of the parameterized circuit, scaling as $\mathcal{O}(\chi_{\rm{ceil}}^3)$ when using TTS, and (ii) quantum tensor programming compilation of the diagonal block encoding, which scales as  $\mathcal{O}(\zeta^3)$ where $\zeta$ is the bond dimension used for the compilation step. In practice, $\zeta\approx \chi_{\rm ceil}=2^{\lceil\log_2\chi_{\rm{base}}\rceil}$ up to small compilation overheads \cite{Termanova2024, Siegl2026}, so 
\begin{equation}
T_{\rm{CL}}\sim (\chi_{\rm{ceil}})^3
\end{equation}
governs the classical overhead \cite{Termanova2024, Siegl2026}.

Because $\succProbs$ and $\epsilon_{\rm{meas}}$ are system-size independent and $d_{\rm{circuit}}$ grows maximally as $\chi^2$, $T_{\rm{CL}}$ becomes the dominant contribution for large bond dimensions. 
Given that three-dimensional turbulent flows require growing bond dimensions with increasing Reynolds number \cite{Gourianov2022}, we expect $T_{\rm CL}$ to be dominant for many industry-relevant CFD regimes.

For comparison, the cost of fully classical tensor network solvers is usually dominated by the evaluation of the nonlinear term, as the deterministic Hadamard product scales as 
$\mathcal{O}(\chi_{\rm TN}^4)$.
While recent algorithmic advances achieve $\mathcal{O}(\chi^3)$ for the Hadamard product via swap-based approaches \cite{Michailidis2025} or the randomized recursive sketch interpolation (RSI) \cite{Meng2026-arxiv}, those often require highly inflated intermediate bond dimensions, reducing significantly the achieved scaling advantage. 
For the RSI, we find significant difficulties in capturing the nonlinearity correctly without strongly increasing the bond dimensions. This leads to a computational cost close to the deterministic computation of the TT-based Hadamard product (cf. Appendix~\ref{ap:RSI-scaling} for the detailed comparison).
The representation difficulties of RSI stem from the bottleneck of the interpolation methods: undersampled points cannot capture the sudden jumps in the field that typically occur in TTs with the rugged underlying data and steep discontinuities \cite{Xiao2026-arxiv}.
\begin{figure}[bt]
    \centering
    \includegraphics[width=0.8\linewidth]{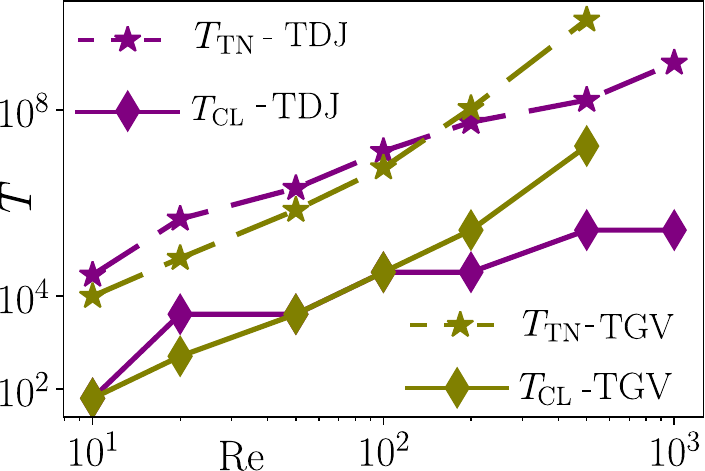}
    \caption{ Computational cost $T_{\rm{TN}}=\chi_{\rm{TN}}^4$ of the fully classical tensor network solver and the classical overhead $T_{\rm CL}=\chi_{\rm ceil}^3$ of the hybrid solver. We use $\chi_{\rm ceil}=2^{\lceil\log_2(\chi_{\rm base})\rceil}$ and the values of $\chi_{\rm{TN}}$ and $\chi_{\rm base}$ shown in \autoref{fig:chi_min}, with $\chi_{\rm base}$ from \textit{TT-Reconstruction + Circuit Approx.}
    For the hybrid scheme, the cost of the classical step  dominates at high bond dimensions as required in highly turbulent regimes.}
	\label{fig:cost-analysis}
\end{figure} 
Hence, here we adopt the scaling with 
\begin{equation}
T_{\rm TN }\sim(\chi_{\rm TN}^4)
\end{equation}
as the practical classical baseline.

Figure~\ref{fig:cost-analysis} compares $T_{\rm{TN}}$ and $T_{\rm CL}$ across Reynolds numbers for both flow configurations.
The hybrid scheme yields a substantial reduction in classical computational overhead up to three orders of magnitudes for the highest Reynolds numbers. Importantly, the ratio $T_{\rm{TN}}/T_{\rm CL}$ enlarges as $\rm Re$ increases indicating the potential for even higher savings for highly turbulent flows.%\nis{not sure if we need to mention industry-relevant in this context, maybe turbulent flows is enough on its own}} 

This scaling analysis assumes comparable training iterations for both approaches. Warm start strategies leverage previous time steps as initial guess and facilitate the training \cite{Siegl2026}. Both frameworks also provide direct access to the fidelity metric 
$\mathcal{F}$, enabling adaptive hyperparameter tuning during optimization. For both approaches the number of qubits and the number of tensors scales logarithmically with $N$.

Beyond computational scaling, the hybrid approach reduces storage requirements for time-dependent fields given the reduced parameter requirements for quantum circuits representing turbulent fields \cite{Siegl2026} and the reduced bond dimension of the TT-reconstruction of $\phi^{[j]}$ compared to the fully classical scheme.

%\textcolor{olive}{I am working with l2 error everywhere. On the quantum circuit I actually measure Fidelity. Do you think this is a problem for this work?}

\section{Conclusion and Outlook} \label{sec:conclusion}
We have presented a hybrid quantum-classical tensor network framework for solving nonlinear partial differential equations in computational fluid dynamics that allows for significant reductions in the computational cost compared to both classical tensor network methods and previous VQA-based routines. %\nis{Provide seems to be overclaim, because we do not show any actual simulation of the eq. with an advantage timing}. 
By combining variational time-stepping with tensor-programming-based operator compilation, the method introduces a novel treatment of nonlinear terms that circumvents the exponentially decaying success probabilities that arise when nonlinearities are implemented via direct point-wise multiplication of quantum states. This stabilization of success probabilities eliminates the prohibitive measurement overhead of previously limited variational quantum CFD algorithms.
Applied to the time evolution of turbulent flows, our hybrid scheme demonstrates stable success probability of $15-30\%$ across increasing Reynolds number and system sizes. 

The additional classical processing step reconstructs an approximate tensor train representation of the velocity field using one of three distinct strategies: two measurement-based protocols and one approach that classically simulates the state encoding quantum circuit using TTs. In all cases, we exploit the fact that high-fidelity state reconstruction is not necessary, as approximate representations of the first nonlinear component suffice for accurate time evolution. This is in contrast to the classical tensor network  simulation where the field differentiation, and the additional contraction and truncation steps lead to significantly higher bond dimension requirements.
With that, the classical reconstruction step offers savings up to three orders of magnitude compared to fully classical tensor network solvers. 
%\nis{this sounds like you actually ran these, in the scaling comparison, I guess also pre-factors need to be considered}. 
As these cost reductions increase with the Reynolds numbers, even higher savings are expected if turbulence increases.
%\nis{do you say "amount of turbulence"?} 
%is further increased. Furthermore, the obtained TT can be used beyond the implementation of the nonlinearity, e.g., for data analysis and the computation of observables \cite{Xiao2026-arxiv}.}

The methodology extends naturally to multi-field %\nis{larger? instead of multi-field}
PDE systems, where the framework is particularly advantageous when simulating coupled fields with disparate %\nis{disparate?} 
compression properties: quantum circuits can efficiently encode high-bond-dimension fields, while classical reconstruction remains efficient for lower-complexity fields. This asymmetry suggests strong potential for problems like Rayleigh–Bénard convection, where temperature demands high bond dimensions \cite{vanHuelst2026RB}. The diagonal block encoding offers the same advantages for scenarios involving the Hadamard product of a time-independent field and a time-dependent solution. Because the corresponding diagonal operator can be precompiled, the algorithm maintains high %\nis{invariant maybe?} 
success probabilities across all time steps without repeated classical reconstruction. Such setups are common in immersed boundary methods with fixed mask functions~\cite{Peddinti2024} and scalar transport in stationary background fields~\cite{Gourianov2024}.

The demonstrated resource efficiency and scalability of this hybrid approach motivate several avenues for further optimization and broader deployment. First, extending the hybrid nonlinear treatment to other classes of PDEs -- including reaction-diffusion systems, magnetohydrodynamics, and multiphase flows -- will clarify the generality of the success probability stabilization. Second, refining the quantum-classical interface through advanced reconstruction protocols and adaptive resource management presents a key opportunity to lower measurement overhead and classical computational costs. Future work should prioritize the systematic evaluation of alternative sampling strategies for the TT-reconstruction in tandem with adaptive bond-dimension control and error mitigation, to quantify their combined influence on shot requirements, reconstruction speed, and numerical accuracy. Finally, hardware-aware porting and optimization as demonstrated in \cite{Over2026-arxiv} will be necessary to translate the algorithmic design into efficient circuits for near-term and fault-tolerant architectures. 
%By reconciling the expressive power of quantum circuits with classical tensor networks, this work establishes a scalable, resource-aware pathway toward practical quantum advantage in turbulent fluid dynamics.

%Finally, coupling the hybrid solver with established CFD preconditioners and domain-decomposition techniques will facilitate integration into existing simulation workflows.

\section*{Acknowledgment}
P.S. acknowledges financial support by the DLR-Quantum-Fellowship Program. P.S. and M.M.B. acknowledge funding from the DLR Quantum Computing Initiative (\url{qci.dlr.de/projects/toquaflics}) and the Federal Ministry for Economic Affairs and Climate Action. 
NL.v.H., T.H., and D.J. are supported by the European Union’s Horizon Europe
research and innovation program (HORIZON-CL4-2021-DIGITAL-EMERGING-02-10) under grant agreement No. 101080085 QCFD.\\
NL.v.H, T.H. and D.J. acknowledge funding from the project ID 390715994 from 16th German-Israeli Call for Proposals for Joint R\&D Projects - Eureka Network.\\
D.J. acknowledges support by the Cluster of Excellence ``Advanced Imaging of Matter'' of the Deutsche Forschungsgemeinschaft (DFG) – EXC 2056 -  and from the Hamburg Quantum Computing Initiative (HQIC) project EFRE. The project is co-financed by ERDF of the European Union and by ``Fonds of the Hamburg Ministry of Science, Research, Equalities and Districts (BWFGB)''.\\

\section*{Data Availability}
The data that support the findings of this article are openly available \cite{dataset}.

\section*{Appendices}
%\onecolumngrid
%\appendix
\appendix
\renewcommand{\thesection}{\Alph{section}}
\renewcommand{\thesubsection}{\Roman{subsection}}

\section{Success Probabilities of the Hadamard Product for Different Fields}\label{ap:randhadamardprod}
In the following, we derive the scaling law of the success probability for performing state-based and block-encoding-based 
Hadamard products of different fields (uniform, uniform-random, and Haar-random) encoded as quantum states.
As discussed in the main text, the optimal success probability of the state-based Hadamard product between fields $\Psi$ and $\phi$
is given by
\begin{align}
   \alpha_{\mathrm{state}}(\phi,\Psi) = \frac{\|D_{\phi}\ket{\Psi}\|_2^2}{\|\phi\|_2^2}
\end{align}
and its block-encoding-based counterpart is given by
\begin{align}
   \alpha_{\mathrm{BE}}(\phi,\Psi)  = \frac{\|D_{\phi}\ket{\Psi}\|_2^2}{\|D_{\phi}\|^2},
\end{align}
where $\ket{\Psi}$ is a quantum state that amplitude encodes $\Psi$ up to its normalization,
$D_{\phi}$ is a diagonal operator that encodes the field $\phi$, and  $\|\phi\|_2$ and $\|D_{\phi}\|$
are the $\ell_2$ norm of a vector and the spectral norm of an operator respectively. 
For simplicity, 
we set $\Psi = \phi$ for the rest of this appendix,
and derive how both success probabilities scale with the system size.

We first derive the scaling for a uniform real field without randomness, 
$$\phi_{\mathrm{uni}}(x_l)=1,$$
defined on $N$ grid points.
The success probability is then given as \begin{equation}
\alpha_{\mathrm{state}}=(\phi_{\mathrm{uni}},\phi_{\mathrm{uni}})=\|\ket{\phi_{\mathrm{uni}}}\|_2^2/N=1/N
\end{equation}
and 
\begin{equation}
\alpha_{\mathrm{BE}}(\phi_{\mathrm{uni}},\phi_{\mathrm{uni}})=\|\ket{\phi_{\mathrm{uni}}}\|_2^2=1,
\end{equation}
using that $\|\ket{\phi_{\mathrm{uni}}}\|_2=1,~ \|\phi_{\rm uni}\|_2=\sqrt{N}$ and $\|D_{\phi_{\rm uni}}\|=1$.

Next we consider the field 
\begin{align}
   \phi_{\mathrm{RU}}(x_l) = u_l
\end{align}
where $u_l$ is drawn from the uniform distribution $\mathcal{U}$ on the interval $[0,1]$. 
In this case, the success probability per realization is given by
\begin{align}
   \alpha_{\mathrm{state}}(\phi_{\mathrm{RU}},\phi_{\mathrm{RU}})
   &=\frac{\sum_l^{N-1} u_l^4}{(\sum_{l}^{N-1}u_l^2)(\sum_{l}^{N-1}u_l^2)}\nonumber\\
   &= \frac{1}{N}\frac{\frac{1}{N}\sum_l^{N-1} u_l^4}{\frac{1}{N}(\sum_{l}^{N-1}u_l^2)\frac{1}{N}(\sum_{l}^{N-1}u_l^2)}\nonumber\\
   &\approx \frac{\mathbb{E}[\mathcal{U}^4]}{N\mathbb{E}[\mathcal{U}^2]^2}
   =\frac{1/5}{N/9} =\frac{9}{5N}
\end{align}
for large $N$, where $\mathbb{E}[\cdots]$ denotes the expectation value, 
and the $k$th-moment of the uniform distribution is given by
\begin{align}
   \mathbb{E}[\mathcal{U}^k] = \int_{0}^{1} x^k  \ dx = \frac{1}{k+1}
\end{align}
Similarly, for block encoding, we obtain, 
\begin{align}
   \alpha_{\mathrm{BE}}(\phi_{\mathrm{RU}},\phi_{\mathrm{RU}})
   &=\frac{\sum_l^{N-1} u_l^4}{(\sum_{l}u_l^2)\max u_l^2}\nonumber\\
   &=\frac{\frac{1}{N}\sum_l^{N-1} u_l^4}{\frac{1}{N}(\sum_{l}^{N-1}u_l^2)\max u_l^2}\nonumber\\
   &\approx \frac{\mathbb{E}[\mathcal{U}^4]}{\mathbb{E}[\mathcal{U}^2]} = \frac{3}{5}. 
\end{align}

\begin{figure}[bt]
    \centering
\includegraphics[width=0.85\linewidth]{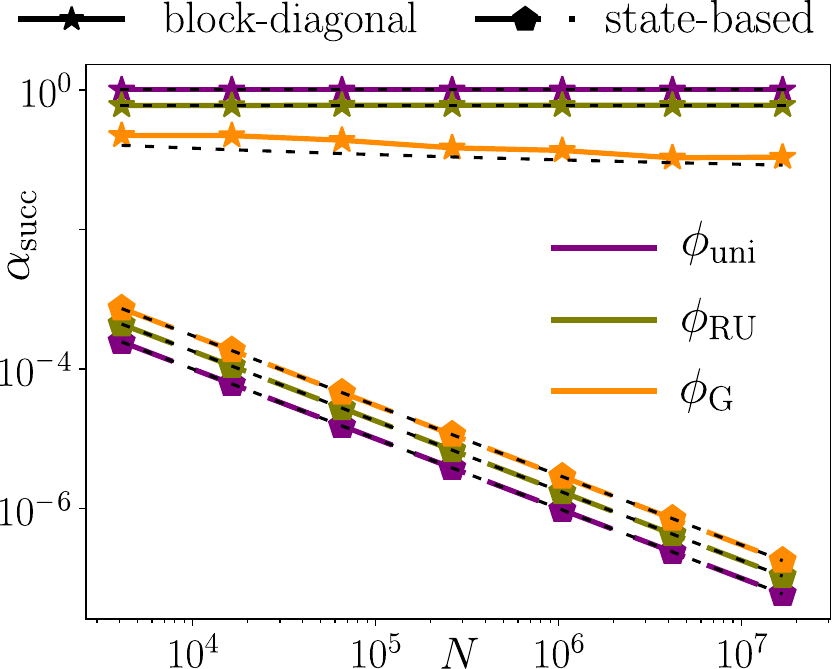}
    \caption{Numerically computed success probabilities $\succProbs$ when computing the square of a field $\phi$ for $\phi$ being the uniform vector $\phi_{\rm uni}$ (purple line), a vector drawn from a uniform distribution $\phi_{\rm RU}$ (olive line) and a vector drawn from the real Haar random distribution $\phi_{\rm G}$  (orange line). We consider $\succProbs$ for the state based approach (pentagons, dashed line) and the block-diagonal based approach (stars, solid line). The theoretical baselines are plotted as black-dotted lines and show excellent agreement. The small deviation for the block-encoded implementation of the real Haar random state, stems from the $\mathcal{O}(1)$ contribution in \autoref{eq:app_alphaHaar} which is neglected here in the baseline computation.
     }
     \label{fig:app_succ_probs}
\end{figure}

Lastly, we show the scaling law for a field whose values $g_l$ at the points $x_l$ are taken from the 
Gaussian distribution $\mathcal{G}$ with mean $0$ and unit variance. 
\begin{align}
   \phi_G(x_l) = g_l
\end{align}
The amplitude encoding of this field, $\ket{\phi_G}$ is equivalent to a state drawn from the real Haar-random distribution, 
which shares the same statistics as the eigenvectors of the Gaussian orthogonal ensemble \cite{Mezzadri2007}.
Similarly to the uniform-random case, we use the well-known expression for the $k$th moment of the Gaussian distrubtion
\begin{align}
   \mathbb{E}[\mathcal{G}^k] = 
   \begin{cases}
       (k-1)!! &(\text{$k$ is even})\\
       0 &(\text{otheriwse})
   \end{cases},
\end{align}
where $(k-1)!!$ is the double factorial, where if $k-1$ is an odd number, it is the product of all the odd numbers starting from $1$ up to and including $k-1$.
With this identity, we obtain
\begin{align}
   \alpha_{\mathrm{state}}(\phi_{\mathrm{G}},\phi_{\mathrm{G}})
   &=\frac{\sum_l^{N-1} g_l^4}{(\sum_{l}^{N-1}g_l^2)(\sum_{l}^{N-1}g_l^2)}\nonumber\\
   &= \frac{1}{N}\frac{\frac{1}{N}\sum_l^{N-1} g_l^4}{\frac{1}{N}(\sum_{l}^{N-1}g_l^2)\frac{1}{N}(\sum_{l}^{N-1}g_l^2)}\nonumber\\
   &\approx \frac{\mathbb{E}[\mathcal{G}^4]}{N\mathbb{E}[\mathcal{G}^2]^2} =\frac{3}{N}.
\end{align}

For the block-encoding-based method, we derive the asymptotic behavior of $\max u_l^2$. 
To derive this, we first find a threshold value for which, on average, there is at least one $g_l^2$ that is larger than $\Theta$.
This is nothing but a tail probability, that of the event $|g_l| > \sqrt{\Theta}$. 
From the cumulative distribution function of the Gaussian, we obtain this probability as
\begin{align}
   P(|g_l| > \sqrt{\Theta}) &= 1 - \mathrm{erf}(\sqrt{\Theta}/\sqrt{2}) \nonumber\\
                            &= 1 - \frac{2}{\sqrt{\pi}}\int_0^{\sqrt{\Theta/2}} e^{-z^2}dz. 
\end{align}
In the limit $1\ll \Theta$, we perform an asymptotic expansion and obtain
\begin{align}
   P(|g_l| > \sqrt{\Theta}) \approx \sqrt{\frac{2}{\pi\Theta}}e^{-\Theta/2}.
\end{align}
The threshold at when the expected number of field values exceeding $\Theta$ equals one is therefore the value near which the maximum is likely to lie. 
Therefore, by solving
\begin{align}
   NP(|g_l| > \sqrt{\Theta}) = 1, 
\end{align}
for $1\ll \Theta$ and $1\ll N$, we obtain the solution $\Theta_c \approx W(2N^2/\pi) = 2\ln N + \ln \ln N + \mathcal{O}(1)$, 
where $W(\cdots)$ is a Lambert $W$ function, and we have used its asymptotic expansion \cite{NIST:DLMF}. 
Altogether, the success probability of the block-encoding method decreases slowly with the number of qubits $n=\log_2 N$
\begin{align}\label{eq:app_alphaHaar}
   \alpha_{\mathrm{BE}}(\phi_{\mathrm{G}},\phi_{\mathrm{G}})
   &=\frac{\sum_l^{N-1} g_l^4}{(\sum_{l}^{N-1}g_l^2) \max g_l^2 } \nonumber\\
   &\approx \frac{3}{2n\ln2+ \ln(n\ln 2) + \mathcal{O}(1)}.
\end{align}

The theoretically derived relations are compared to the numerically computed success probabilities in \autoref{fig:app_succ_probs} and show excellent agreement.
\section{Implementation Details on Tensor Train Reconstruction Techniques} \label{app:tt-reconstruction}
 Here, we provide the implementation details on the different approaches for the TT-reconstruction step.
 \subsection{Tensor Cross Interpolation} 
 We employed the open-source Julia package \textit{TensorCrossInterpolation.jl} \cite{TCIPaper}. The algorithm constructs a low-rank TT approximation by querying state amplitudes on a sparse, cross-shaped grid subset. For the TT-reconstruction we have used the default TCI settings but have increased the value of \textit{maxiter} to $100$, and the values of  \textit{maxnglobalpivot}, \textit{nsearchglobalpivot} to $50$ as recommended by \cite{TCIPaper}.
 Furthermore we have set a bond dimension $\chi$ and an allowed tolerance. We have found that larger tolerances of size $10^{-4}$ facilitate the search of the minimal required  $\chi_{\rm TCI}$. 
 We have further tried to optimize the TCI by starting with a good initial guess or by using optimization techniques for identify good initial pivots as provided in \textit{TensorCrossInterpolation.jl}. 
 None of these changes had significant impact on the required bond dimension of the target TT to represent the flows and were hence not used for the final data generation. 
To estimate the required bond dimension for the TCI, we have subsequently increased $\chi_{\rm TCI}$ until the target accuracy was reached.
In general, once, a sufficiently good TT-approximation is found with TCI, the resulting TT can be truncated using SVD to a lower bond dimension. 
 %\textcolor{blue}{check!}
 Because TCI relies on stochastic amplitude queries, we repeated the reconstruction 100 times and selected $\chi_{\rm TCI}$ based on the mean relative $\ell_2$-error. At this bond dimension, the variance of $\epsilon_{\ell_2}$ across independent runs remained below $10^{-7}$, indicating robust convergence of the TCI.
To emulate finite-shot measurement noise, we treated each exact amplitude as an expectation value and sampled it probabilistically from a binary distribution, consistent with standard quantum measurement statistics.
%. 
To create the data corresponding to \autoref{fig:mps-tomo}~(b), we have repeated the finite-shot run $10$ times and have considered the mean  of $\epsilon_{\ell_2}$ as the final error.
\subsection{Efficient State Tomography}
The EST protocol proposed in \cite{Cramer2010} reconstructs the tensor train by sequentially measuring reduced density matrices on qubit subsets of size $k$, where the subset size satisfies $2^{k-1} \geq \chi_{\rm base}$. For each subset, a disentangling unitary is derived from the measured density matrix and applied to isolate the corresponding TT core. The accumulated unitaries are then composed to assemble the full TT. We implemented this protocol in Python using the PennyLane framework \cite{pennylane}.
To emulate finite-shot measurement noise, the density matrix estimation was performed via Pauli measurements followed by maximum-likelihood reconstruction. 
For the density matrix estimation, we have utilized threshold quantum state tomography (tQST), a method developed by Binosi et
al. \cite{tQST_Binosi}. In principle, it can 
reduce the number of total measurements,
by measuring only relevant off-diagonal entries determined by 
a given threshold value. In this work, we have set a fixed threshold of $10^{-18}$ for the computations.
As CFD data is intrinsically real, we here consider full quantum state tomography for real-valued density matrices, which reduces the number of measurement shots by nearly a factor $1/2$ without loss of fidelity. To this aim we have adapted the code tailored for complex density matrices provided in \cite{tQST_Binosi_Code}. 
In the end, a maximum likelihood estimation is performed, for which we use a real-triangular model which further reduces the number of parameters optimized over, compared to alternative complex models.
\subsection{Direct Tensor Train Simulation}
To estimate the bond dimension required for the TTS of the field-encoding circuit, we first construct a circuit with realistic depth. Since TT representations provide a direct compilation strategy and a rigorous upper bound on circuit depth, the TT-derived circuit serves as an ideal benchmark for evaluating simulation requirements. We decompose the target field into a TT and compile its cores into unitary gates following the strategy of \cite{Lubasch2020}. The resulting circuit is then decomposed into one- and two-qubit gates using the PennyLane framework \cite{pennylane}. The simulation is performed with PennyLane's \texttt{default.tensor} backend, enforcing a maximum bond dimension $\chi_{\rm ceil}$ during state propagation. We verify that simulating the decomposed circuit of a TT of bond dimension $\chi_{\rm ceil}$ faithfully reproduces the target field accuracy, confirming that the gate decomposition and truncation pipeline preserves the desired representation quality.

\section{Turbulent Data and Simulations} \label{ap:turbDat}
In the following we provide details on the two turbulent flow examples used in the main manuscript.
For the simulation of the evolved field, we have utilized the direct numerical simulation solver used in \cite{Gourianov2022} and provided in \cite{Gourianov2022Code}.
The simulation utilizes a penalty method for the pressure correction and a Runge-Kutta 2 time-stepping routine. We consider the solution after $2000$ time steps to perform the analysis on the developed turbulent field. To ensure a divergence-free solution, we perform an additional pressure correction step to the final field prior to the analysis.
For the temporally developing jet (TDJ), we have considered the Reynolds numbers $\rm{Re}\in [10,20,50,100,200,500,1000]$ and corresponding grid sizes $(2^{n_d})^2$, with $n_d=[6,7,8,9,9,9,10]$.
For the Taylor Green vortex (TGV) we have consider the Reynolds numbers $\rm{Re}\in [10,20,50,100,200,500]$ solved on a grid of size $(2^{n_d})^3$, with $n_d=[4,5,6,6,7,8]$.
For the computation of the success probability we have chosen the time step size $\Delta t$ to satisfy $\frac{\max(|\bm{u}|)\Delta t}{dx}=0.5$.
Furthermore, we have used the viscosities $\nu=1/(200 \textrm{ Re})$ and $\nu=1/(2\pi \textrm{ Re})$ for the TDJ and the TGV respectively, in agreement with \cite{Gourianov2022, Gourianov2022Code}. All spatial derivatives were implemented using central finite differences.

\section{Errors of Fields and Derivatives} \label{app:bond-dim-scaling}
\begin{figure}[bt]
    \centering
\includegraphics[width=\linewidth]{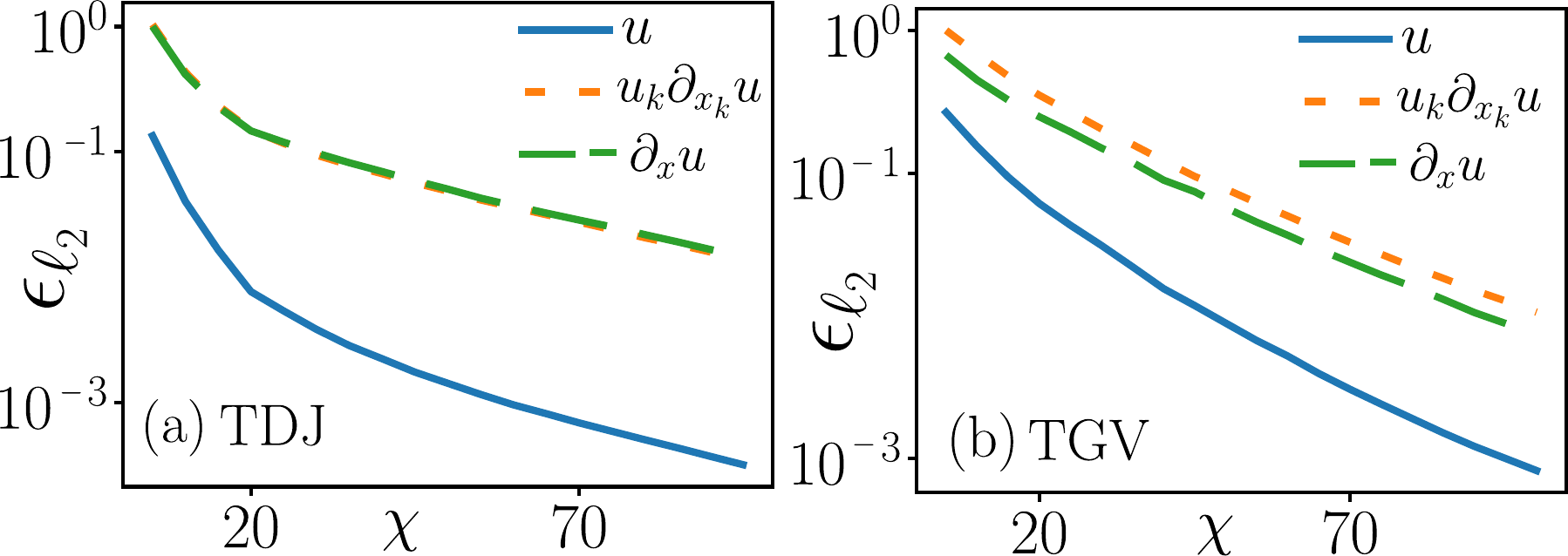}
    \caption{Realitive $\ell_2$-errors $\epsilon_{\ell_2}$ over the bond dimension $\chi$ for the velocity field (blue solid), its spatial derivative (green, dashed) and the nonlinear advection term (orange, dotted). 
    All terms are evaluated deterministically using classical TTs. We focus on the first velocity component $u_1=u$, its derivative $\partial_{x_1}=\partial_x$ and the corresponding nonlinear contribution $A_1 = \sum_{k} u_k\partial _{x_k} u$. Panels (a) and (b) show results for the TDJ  at $\textrm{Re}=500$ and the TGV at $\textrm{Re}=200$, respectively.}
\label{fig:app_eps_over_chi}
\end{figure}
To understand why the hybrid approach tolerates reduced bond dimensions over the classical TT scheme, we analyze how the relative $\ell_2$-error $\epsilon_{\ell_2}$ (defined in \autoref{eq:l2-error})  of individual terms scales with the bond dimension $\chi$.
Figure~\ref{fig:app_eps_over_chi} depicts $\epsilon_{\ell_2}$ for the field, the derivative and the nonlinear term of the TDJ and the TGV flows at two Reynolds numbers. In both cases, the field representation converges significantly faster than its derivative or the full nonlinear term, yielding errors that are one to two orders of magnitude lower for the same $\chi$. The gap between field and derivative errors widens at higher $\chi$, indicating that the compression advantage of the hybrid approach is more pronounced as stricter accuracy targets are imposed.
\section{Error Propagation and Measurement Requirements} \label{ap:error-prop}
In the adapted Hadamard test, the expectation value $\expVal$ approaches unity at optimal training. The standard Monte Carlo error estimate $\epsilon_{\rm MC} = \sqrt{1-\expVal\ell_2}/\sqrt{m}$ vanishes in this limit, suggesting only a single shot is required. In practice, however, $\expVal = 1 - \epsilon_{\expVal}$, where $\epsilon_{\expVal}$ represents the deviation from unity due to finite training precision, optimization tolerance or representing our limited knowledge of its optimality. Setting the target sampling accuracy to $\epsilon^t_{\expVal} \approx \epsilon_{\expVal}$, the required shot count scales as
\begin{equation}
\begin{aligned}
m &= \frac{(1-\expVal)^2}{(\epsilon^t_{\expVal})^2}=  \frac{(1-\expVal)(1+\expVal)}{(\epsilon^t_{\expVal})^2}\\
&\approx\frac{2\epsilon^t_{\expVal}}{(\epsilon^t_{\expVal})^2}=\frac{2}{\epsilon^t_{\expVal}},
\label{eq:shot_reqs_aH}
\end{aligned}
\end{equation}
suggesting a $1/\epsilon^t_{\expVal}$ scaling of the required measurement shots.
However, the algorithm also requires accurate estimation of the normalization correction factor $f_{\hat{O}}$, which depends on the Hadamard test parameter $\varphi$. The precision of $f_{\hat{O}}$ directly dictates the required sampling accuracy. We relate the expectation value error to the parameter error $\epsilon_\varphi$ via a second-order Taylor expansion around the optimal angle $\varphi^{*}$:
\begin{equation}
\epsilon^t_{\expVal} \approx \frac{1}{2} \left| \frac{d^2\expVal}{d\varphi^2} \right|_{\varphi^{*}} \epsilon_\varphi^2,
\label{eq:taylor_exp}
\end{equation}
where we used $\partial_\varphi \expVal|_{\varphi^{*}} = 0$ and assume $\varphi$ is optimized during training. The resulting error in the normalization factor follows from first-order error propagation:
\begin{equation}
\epsilon_{f_{\hat{O}}} \approx \left| \frac{\partial f_{\hat{O}}}{\partial \varphi} \right|_{\varphi^{*}} \epsilon_\varphi.
\label{eq:eps_fo}
\end{equation}
The target accuracy refers to the relative error in the normalization factor, $\epsilon_{\rm meas} = \epsilon_{f_{\hat{O}}} / f_{\hat{O}}$. Substituting $\epsilon_{f_{\hat{O}}} = f_{\hat{O}} \cdot \epsilon_{\rm meas}$ into the error propagation chain 
and combining \autoref{eq:shot_reqs_aH}, \autoref{eq:taylor_exp}, and \autoref{eq:eps_fo} with the probabilistic success probability $\alpha_{\rm succ}$, we obtain the total measurement scaling:
\begin{equation}
m = \mathcal{O}\left( \frac{1}{\alpha_{\rm succ}( \epsilon_{\rm meas})^2}\right).
\end{equation}
This recovers the standard inverse-quadratic scaling of the Hadamard test with respect to the target accuracy.
\section{Scaling Analysis with Recursive Sketch Interpolation} \label{ap:RSI-scaling}
\begin{figure}[bt]
    \centering
\includegraphics[width=\linewidth]{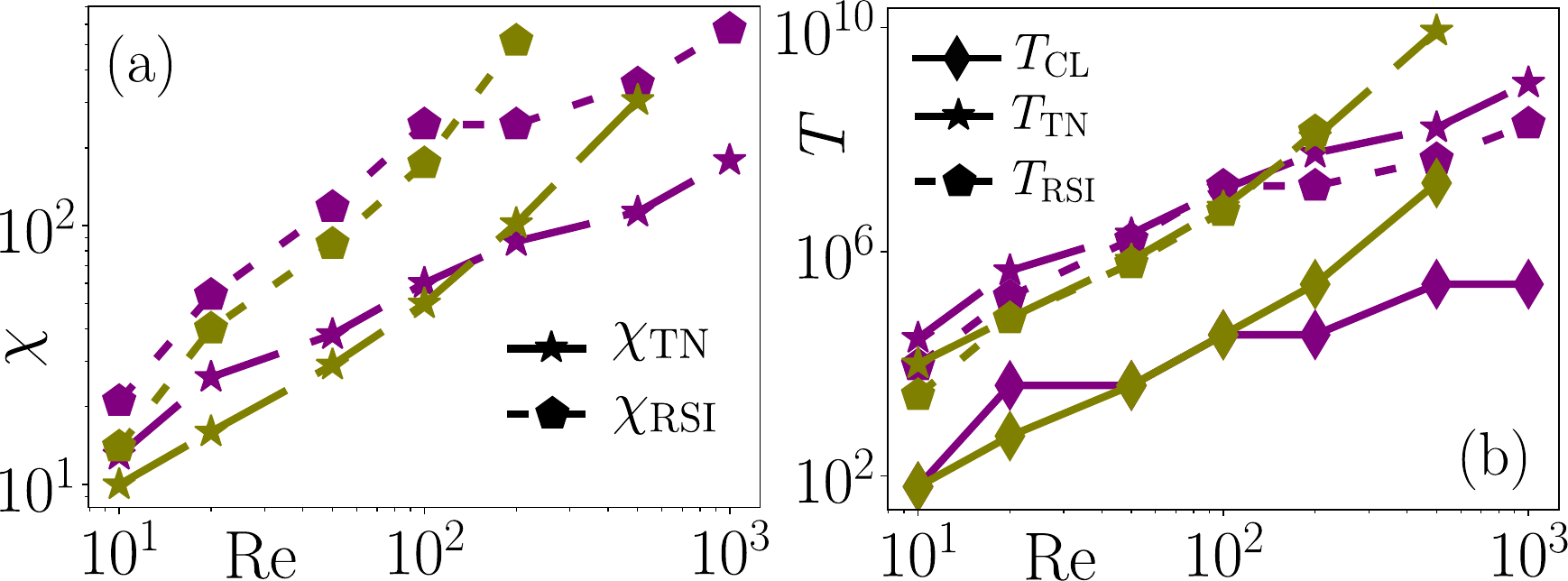}
    \caption{Comparison between the standard computation and the recursive sketch interpolation (RSI) of the nonlinear term $A_i$ for the temporally developing jet (purple lines) and the Taylor Green vortex (olive lines). (a) Bond dimension required to compute all nonlinear advection terms $A_i$ up to a maximal error of $\epsilon_{\ell_2}\leq0.01$ and (b) the resulting computational cost $T$. For the standard TT computation we consider a cost of $T\sim(\chi_{\rm TN}^4)$, while for the RSI base method we consider $T\sim (\chi_{\rm RSI}^3)$. We observed substantial savings in the bond dimension when using the standard approach, while the cost of both methods scales comparatively. For comparison, the classical cost $T_{\rm CL}$ of the hybrid quantum classical approach is shown, being orders of $2-3$ magnitudes below the classical tensor network scheme.
    For the Taylor Green vortex at $\rm{Re}=500$ the RSI has not found a solution within $\chi_{\rm RSI}\leq 1000$. At $\chi_{\rm RSI}=1000$, we still get $\epsilon_{\ell_2}=0.472$, indicating the need for much higher bond dimensions or general problems of convergence.}
     \label{ap:rsi}
\end{figure}
Here, we analyze the computation of the nonlinear terms $A_i$ in the TT format, and
compare two implementations in terms of the required bond dimension $\chi$. First, we consider the standard variational method to compute the TT Hadamard product \cite{Schollwöck2011, Gourianov2022}: %(cite?) 
one factor is promoted to a TTO via the
delta-tensor formalism, and the solution of the TTO--TT contraction is obtained by optimizing a target TT with bond dimension $\chi_{\rm TN}$. This construction is $\ell_2$-optimal at cost $O(\chi_{\rm TN}^4)$ \cite{Schollwöck2011} and yields the same result as a direct TTO--TT contraction followed by truncation to $\chi_{\rm TN}$, up to the optimization tolerance. The second implementation is Recursive Sketched
Interpolation (RSI)~\cite{Meng2026-arxiv} which combines randomized TT sketching with an
interpolative decomposition, constructing the result at cost
$T_{\rm RSI}= O(\chi_{\rm RSI}^3)$ with the output bond dimension denoted by $\chi_{\rm RSI}$.
For the computation with RSI, we use the baseline parameters of Ref.~\cite{Meng2026-arxiv}:
a two-core contraction step ($\mathrm{ccn}=2$), an oversampling parameter
$p=5$, an internal tolerance of $10^{-10}$, and pivoted QR
for the interpolative decomposition. In both approaches, the same maximal
bond dimension $\chi$ is prescribed for the input velocity fields, derivative
TTs, nonlinear products, and final TT sums.

For each velocity component $i$, the nonlinear term $A_i$
(cf.~\autoref{eq:adv-term}) is compared with the dense reference using the
relative $\ell_2$-error $\epsilon_{\ell_2, i}$ defined in
\autoref{eq:l2-error}. For the standard variational TT-method, the required
bond dimension is the smallest $\chi$ satisfying
\[
    \max_i \epsilon_{\ell_2,i}(\chi) < 10^{-2}.
\]
For RSI, ten fixed random seeds are used. We require that for all seeds
\[
    \max_{\substack{i\\ j=0,\ldots,3}}
    \epsilon_{\ell_2,i}(\chi+j) < 10^{-2},
\]
i.e., all components and seeds must satisfy the threshold for four consecutive
bond dimensions, to account for stochastic variations within the RSI.

Figure~\ref{ap:rsi}~(a) shows the smallest bond dimensions satisfying these
criteria. The Hadamard product computed with the standard variational approach requires significantly smaller bond dimensions than RSI and represents the optimally rounded TT result.
In Fig.~\ref{ap:rsi}~(b), these bond dimensions are used to compute the computational costs using the
$O(\chi_{\rm TN}^4)$ scaling of the variational Hadamard-product method and
the $O(\chi_{\rm RSI}^3)$ scaling of RSI. Because RSI requires larger bond
dimensions to obtain the target accuracy, the effective costs become comparable. We therefore retain the $O(\chi_{\rm TN}^4)$ variational TT baseline in the complexity
analysis of the main text.
\FloatBarrier
\bibliography{bibliography.bib}

\end{document}